\documentclass[twocolumn,showpacs,superscriptaddress,floatfix,pra]{revtex4-2}
\usepackage{graphicx,amsfonts,amssymb,amsmath,enumerate}
\usepackage{braket}
\usepackage{graphicx}
\usepackage{dcolumn}
\usepackage{bm}
\usepackage{color}
\usepackage[colorlinks,allcolors=blue]{hyperref}
\usepackage{physics}
\usepackage{adjustbox}
\usepackage{graphicx}
\usepackage{subfigure}
\begin{document}

\preprint{APS/123-QED}

\title{Lattice-induced wavefunction effects on trapped superfluids} 

\author{Yeyang Zhang}
\email{yeyzhang@pku.edu.cn}
\affiliation{International Center for Quantum Materials, School of Physics, Peking 
 University, Beijing 100871, China}


\date{\today}

\begin{abstract}
Wavefunction effects in uncorrelated systems are characterized by the Berry curvature and quantum metric. Beyond those, we propose gauge-independent tensors describing Bloch wavefunction effects on local interaction between correlated particles. We derive an effective hydrodynamic theory for ultracold bosons in optical lattices. Ground states and collective modes of superfluids in isotropic harmonic traps are solved for highly symmetric lattices. In a dynamic process, the wavefunction effects are featured by the eigenfrequency, amplitude, and phase shift of an excited breathing mode and can be observed in experiments. We also give a tight-binding model of a bipartite square lattice with nontrivial wavefunction effects, where results are estimated with typical experimental parameters. Our discovery advances the connections between the modern band theory and quantum many-body physics.
\end{abstract}

\maketitle


\section{\label{Sec1}Introduction}

Quantum systems are traditionally featured by the eigenenergies of Hamiltonians. Lattices modify the low-energy spectra of electrons or atoms through effective mass~\cite{Luttinger1955, Kittle2004}. Eigenstates of quantum systems are also crucial for physical observables. As global properties of Bloch wavefunctions, topological indices distinguish insulator phases~\cite{Schnyder2008, Schnyder2009, Kitaev2009, Ryu2010, Gong2018, Kawabata2019} and lead to quantized Hall conductances~\cite{Klitzing1980, Laughlin1981, Thouless1982, Haldane1988, Chang2013, Kane2005, Kane2005_2, Bernevig2005, Bernevig2006, Konig2007}. Local properties of wavefunctions are quantified by the quantum geometric tensor, i.e.~the Berry curvature and quantum metric~\cite{Pancharatnam1956, Provost1980, Berry1984}. Quantum geometric effects on linear~\cite{Karplus1954, Chang1996, Ganesh1999, Nagaosa2010, Xiao2010} and nonlinear~\cite{Gao2014, Du2021, Du2021_2, Wang2021, Wang2023, Gao2023} transport have been widely studied.

Wavefunction effects of single-particle Bloch states play important roles even in many-body systems. It indicates possible fractional Chern insulators~\cite{Parameswaran2012, Roy2014, Jackson2015} and stabilizes superfluid~\cite{You2012, Julku2021, Julku2021_2, Peotta2023} or superconductor~\cite{Peotta2015, Julku2016, Liang2017, Iskin2020, Wang2020, Torma2022, Kitamura2022, Iskin2023, Chen2023, Hu2023, Tian2023} phases in flat-band systems. For trapped interacting bosonic atoms in optical lattices~\cite{Kramer2002, Cataliotti2003, Morsch2006, Konabe2007, Hu2010}, the wavefunction effects on hydrodynamic equations have been proposed based on the Berry curvature~\cite{Price2013, Li2017}. However, the spatial density fluctuation of atoms is comparable with the lattice for a ground state, so it is improper to treat the interaction as a slowly varying mean-field potential as in the literature. Matrix elements between two-body Bloch states are necessary to characterize the correlation between particles. Despite some results of Bogoliubov excitation spectra~\cite{Subasi2022, Lukin2023}, a hydrodynamic theory extracting the wavefunction effects beyond the single-body quantum geometric tensor is still lacking.

In this paper, we derive a low-energy effective hydrodynamic theory of locally interacting bosons in an optical lattice with an additional external potential. The spatial variation of the external potential is assumed to be small compared with the lattice, so a generalized effective mass theory with a gradient expansion~\cite{Li2018} applies to the system. Distinct from the single-body external potential, in the second order of momentum, a $\phi^4$ interaction is corrected by one scalar, one vector, and three second-order symmetric tensors. Those quantities characterize \textit{lattice-induced (Bloch) wavefunction effects} of the system. All of the quantities are invariant under a gauge transformation in the momentum space. One of the second-order tensors is related to inverse participation ratios~\cite{Evers2008} of Bloch wavefunctions in a unit cell. Another is odd under time reversal, whose symmetry is forbidden in the quantum geometric tensor. For a three-dimensional (3D) lattice of the cubic crystal system or a two-dimensional (2D) lattice with a $\mathbb{Z}_n$ ($n\geq 3$) rotational symmetry, the effective theory becomes isotropic when the external potential is central. Without time-reversal symmetry, the Thomas-Fermi distribution~\cite{Li2017, Zhai2021} for a ground state is corrected with a finite spatial variation of the phase of bosons.

Our corrections to the hydrodynamic theory stem from a different physical origin, compared with those due to quantum fluctuations, e.g.~the Lee-Huang-Yang correction~\cite{Pitaevskii1998, Braaten1999}. Although both of them are many-body effects, the lattice-induced wavefunction effects are already displayed in the classical limit of the boson field. In a weakly interacting limit in 2D or 3D, where interaction becomes small enough while chemical potential keeps in the same order, the quantum fluctuation corrections are negligible compared with the lattice corrections. The classical hydrodynamic description is valid in this limit. The results in this paper are derived therein and can be generalized with the quantum fluctuations.

For measurable predictions, we solve collective modes in a harmonic trap~\cite{Stringari1996, Perez1996, Dalfovo1999, Onofrio2000} for the isotropic cases. The ground state depends on the lattice potential, so the collective modes can be excited and observed by suddenly varying the optical lattice at some time. In the leading order of the gradient expansion, values of the scalar and two of the three second-order tensors describing the lattice-induced wavefunction effects can be differentiated experimentally. When the wavefunction effects are weak, they are directly linked to three observables, the frequency, amplitude, and initial phase of a breathing mode. To make a nonzero initial phase, the time-reversal symmetry needs to be broken. As a concrete example of tight-binding models, we consider a bilayer square lattice in a nonuniform synthetic magnetic field~\cite{Ho1996, Lin2009, Zheng2014}, whose Bloch wavefunction effects are only exhibited with interaction. Therein, the main results of this paper are estimated with typical orders of magnitude of experimental parameters~\cite{Bloch2005, Jo2012, Tarruell2012, Jotzu2014, Taie2015, Aycock2017}. Although the lattice model is a theoretical prototype, essential physics is shown, and our framework applies to general superfluids in lattices.

This paper is structured as follows. In Sec.~\ref{Sec2}, we derive the continuous-space effective theory of superfluids in lattices with $\phi^4$ interaction and show the lattice-induced effective corrections to the interaction. In Sec.~\ref{Sec3}, we calculate the hydrodynamic properties of the superfluids based on the effective theory, including ground states and collective modes of the system. We also discuss when the quantum fluctuations are negligible compared with the lattice-induced corrections. In Sec.~\ref{Sec4}, we apply the hydrodynamic theory where the external potential is the isotropic harmonic trap. In particular, we propose the direct observables of the lattice-induced wavefunction effects. In Sec.~\ref{Sec5}, we provide the tight-binding example with nontrivial wavefunction effects. A brief summary and outlook are given in Sec.~\ref{Sec6}. Technical details for deriving the effective interaction and solving the collective modes in the harmonic trap are shown in Appendix \ref{AppendixA} and Appendix \ref{AppendixB}, respectively. Further discussions on tight-binding modes are offered in Appendix \ref{AppendixC}.

\section{\label{Sec2}Effective theory of superfluids in lattices}

We start from a continuous model of bosonic atoms in $d$ dimensions. The Hamiltonian of the system is given by three parts, $H=H_0+H_U+H_I$. Here $H_0$ is a single-body Hamiltonian with an optical lattice, and it is diagonalized in a Bloch basis ($\hbar=1$),
\begin{align}
\label{eqn1}
H_0=\sum_{\mathsf{n}}\int_{\mathrm{BZ}} d^d\bm{k}b^\dagger_{\mathsf{n},\bm{k}}\epsilon_{\mathsf{n}}(\bm{k})b_{\mathsf{n},\bm{k}},
\end{align}
where $\mathsf{n}$ is a band index and ``BZ" denotes an integral in the first Brillouin zone. The bottom of the lowest band ($\mathsf{n}=0$) is assumed at $\bm{k}=\bm{0}$, around which there is a quadratic dispersion, $\epsilon_{0}(\bm{k})\approx\frac{k^2}{2m}$. Bosonic operators in the real space and Bloch space are linked by Bloch wavefunctions,
\begin{align}
\label{eqn2}
a(\bm{r})=\sum_{\mathsf{n}}\int_{\mathrm{BZ}}d^d\bm{k}e^{i\bm{k}\cdot\bm{r}}u_{\mathsf{n},\bm{k}}(\bm{r})b_{\mathsf{n},\bm{k}}.
\end{align}
$u_{n,\bm{k}}(\bm{r})$ is normalized such that the average of $|u_{0,\bm{k}}(\bm{r})|^2$ in a unit cell equals $(2\pi)^{-d}$. $H_U$ is an applied external potential and $H_I$ is a local repulsive interaction between atoms,
\begin{align}
\label{eqn3}
H_U=&\int_\infty d^d\bm{r} a^\dagger(\bm{r})U(\bm{r})a(\bm{r}),\nonumber\\
H_I=&\frac{g}{2}\int_\infty d^d\bm{r}a^\dagger(\bm{r})a^\dagger(\bm{r})a(\bm{r})a(\bm{r}).
\end{align}

In cold-atom systems, it is proper to model the interaction to be short-ranged~\cite{Zhai2021, Sadeghpour2000, Chin2010} even at the length scale of the optical lattice. The spatial variation of $U(\bm{r})$ is assumed to be small compared with the lattice, such that the Fourier components of $U(\bm{r})$ outside the first Brillouin zone are negligible~\cite{Li2018}. Besides, suppose only the excitation of Bloch states near the band bottom of the lowest band are considered in a low-energy effective theory, so the position index $\bm{r}$ of a Wannier basis $b_{\mathsf{n}}(\bm{r})$ (i.e.~a Fourier transform of the Bloch basis) is approximated to be continuous in an effective theory. Those assumptions require that the typical energy and momentum scales of the external potential are much smaller than those of the chemical potential and band dispersion. Those assumptions apply to ground states and low-energy hydrodynamic collective modes with slowly varying $U(\bm{r})$ (see Secs.~\ref{Sec3} and \ref{Sec4}).

Taking hydrodynamic variables, i.e.~$b(\bm{r})=\sqrt{n(\bm{r})}e^{i\theta(\bm{r})}$, and letting the field variables depend on time, we get an effective action in the (coarse-grained) Wannier basis by a second-order expansion of the (quasi)momentum $\bm{k}$,
\begin{align}
\label{eqn4}
\mathcal{S}=\int dt[\int_\infty d^d\bm{r}(-n\partial_t \theta+\mu n)-H_0-H_U-H_I],
\end{align}
where the theory has been projected to the lowest band (the band index is omitted) and $\mu$ is a chemical potential. The free Hamiltonian of quasiparticles reads
\begin{align}
\label{eqn5}
H_0=\int_\infty d^d\bm{r}\frac{1}{2m}\big{[}\frac{(\bm{\nabla}n)^2}{4n}+n(\bm{\nabla}\theta)^2\big{]}.
\end{align}

\subsection{\label{Sec21}Effective external potential}

The effective correction to the external potential by the Berry curvature ($\Omega_\gamma$) and quantum metric ($g_{\alpha\beta}$) have been derived in the literature~\cite{Li2018},
\begin{align}
\label{eqn6}
H_U=\int_\infty d^d\bm{r}&\big{[}Un+\frac{1}{2}\mathsf{g}_{\alpha\beta}(\bm{0})n\partial_\alpha\partial_\beta U\nonumber\\
&-\frac{1}{2}\bm{\Omega}(\bm{0})\cdot(\bm{\nabla}U\times\bm{\nabla}\theta)n\big{]},
\end{align}
\begin{align}
\label{eqn7}
\Omega_\gamma(\bm{k})&=\frac{1}{2}\epsilon_{\alpha\beta\gamma}\Omega_{\alpha\beta}=\frac{1}{2}\epsilon_{\alpha\beta\gamma}[\partial_{k_\alpha} A_\beta(\bm{k})-\partial_{k_\beta} A_\alpha(\bm{k})],\nonumber\\
\mathsf{g}_{\alpha\beta}(\bm{k})&=\frac{1}{2}[\langle\partial_{k_\alpha} u_{\bm{k}}|(1-|u_{\bm{k}}\rangle\langle u_{\bm{k}}|)|\partial_{k_\beta} u_{\bm{k}}\rangle+(\alpha\leftrightarrow\beta)],\nonumber\\
A_\alpha(\bm{k})&=i\langle u_{\bm{k}}|\partial_{k_\alpha} u_{\bm{k}}\rangle.
\end{align}
A gauge transformation $u_{\bm{k}}\rightarrow u_{\bm{k}}e^{i\vartheta_{\bm{k}}}$ has been applied when deriving Eq.~(\ref{eqn6}) such that the result becomes ``gauge-independent",
\begin{align}
\label{eqn8}
\vartheta_{\bm{k}}=A_\alpha(\bm{0})k_\alpha+\frac{1}{4}[\partial_{k_\alpha} A_\beta(\bm{0})+\partial_{k_\beta} A_\alpha(\bm{0})]k_\alpha k_\beta.
\end{align}
Actually, the result is gauge-dependent, while it equals a gauge-independent quantity in the specific gauge.

Note that $n(\bm{r})$ in Eq.~(\ref{eqn4}) also depends on the gauge. As an observable, the (gauge-independent) physical density of the atoms $n_{\mathrm{ph}}(\bm{r})$ should be derived from the physical external potential $U(\bm{r})$,
\begin{align}
\label{eqn9}
&n_{\mathrm{ph}}(\bm{r})=-\frac{\delta\mathcal{S}}{\delta U(\bm{r})}\nonumber\\
=&n+(\bm{\nabla} n)\cdot(\bm{\nabla}\theta\times\bm{\Omega})+\frac{1}{2}\mathsf{g}_{\alpha\beta}\partial_\alpha\partial_\beta n.
\end{align}
Here the time argument is omitted and $U(\bm{r,t})=U(\bm{r})$. The physical density can also be obtained when counting the chemical potential $\mu$ into the external potential $U(\bm{r})$. Because $\mu$ is a constant, from Eq.~(\ref{eqn6}) we know the result Eq.~(\ref{eqn9}) is unaffected by such substitution.

\subsection{\label{Sec22}Effective interaction}

To derive an effective interaction, we write the $\phi^4$ interaction in the Bloch basis,
\begin{align}
\label{eqn10}
\frac{2}{g}H_I=\frac{1}{(2\pi)^d}\int_{\mathrm{BZ}}(\prod_{i=1}^3 d^d\bm{k}_i)b^\dagger_{\bm{k}_1}b^\dagger_{\bm{k}_2}b_{\bm{k}_3}b_{\bm{k}_4}\langle\langle u_{\bm{k}_1}u_{\bm{k}_2}|u_{\bm{k}_3}u_{\bm{k}_4}\rangle\rangle,
\end{align}
where $\bm{k}_4\equiv\bm{k}_1+\bm{k}_2-\bm{k}_3$,
\begin{align}
\label{eqn11}
&\langle\langle u_{\bm{k}_1}u_{\bm{k}_2}|u_{\bm{k}_3}u_{\bm{k}_4}\rangle\rangle\nonumber\\
\equiv&\frac{(2\pi)^{2d}}{\Omega_{\mathrm{cell}}}\int_{\mathrm{cell}}d^d\bm{r}u^*_{\bm{k}_1}(\bm{r})u^*_{\bm{k}_2}(\bm{r})u_{\bm{k}_3}(\bm{r})u_{\bm{k}_4}(\bm{r}),
\end{align}
``$\mathrm{cell}$" denotes an integral in the unit cell, $\Omega_{\mathrm{cell}}$ is the volume of the unit cell. A relation has been applied when taking a sum over lattice vectors $\bm{R}$,
\begin{align}
\label{eqn12}
\sum_{\bm{R}}e^{i(\bm{k}_3+\bm{k}_4-\bm{k}_1-\bm{k}_2)\cdot\bm{R}}=\frac{(2\pi)^d}{\Omega_{\mathrm{cell}}}\delta^d(\bm{k}_1+\bm{k}_2-\bm{k}_3-\bm{k}_4).
\end{align}
Only half of the first Brillouin zone near the band bottom has been considered such that reciprocal scattering has been neglected. We further approximate the integral domain of momenta in Eq.~(\ref{eqn10}) to be infinity to get an effective theory in the continuous space.

Then we can apply a second-order expansion of momentum. We get a zeroth-order term~\cite{Kramer2002},
\begin{align}
\label{eqn13}
H_{I0}=\frac{g}{2}\int_\infty d^d\bm{r}M(\bm{0})n(\bm{r})^2,
\end{align}
a first-order term,
\begin{align}
\label{eqn14}
H_{I1}=\frac{g}{2}\int_\infty d^d\bm{r}M'_\alpha(\bm{0})n^2(\partial_\alpha\theta),
\end{align}
and a second-order term (see Appendix \ref{AppendixA1}),
\begin{align}
\label{eqn15}
&H_{I2}=\frac{g}{2} \int_\infty d^d\bm{r}\big{[}\frac{1}{2}M''_{\alpha\beta}(\bm{0})n^2(\partial_\alpha\theta)(\partial_\beta\theta)\nonumber\\
&\!\ \!\ -2\tilde{S}_{\alpha\beta}(\bm{0})n(\partial_\alpha n)(\partial_\beta\theta)+W_{\alpha\beta}(\bm{0})(\partial_\alpha n)(\partial_\beta n)\big{]},
\end{align}
where
\begin{align}
\label{eqn16}
&M(\bm{k})=\langle\langle u_{\bm{k}}u_{\bm{k}}|u_{\bm{k}} u_{\bm{k}}\rangle\rangle\nonumber\\
=&1+\frac{(2\pi)^{2d}}{\Omega_{\mathrm{cell}}}\int_{\mathrm{cell}}d^d\bm{r}\big{[}|u_{\bm{k}}(\bm{r})|^2-\frac{1}{(2\pi)^d}\big{]}^2\geq 1,
\end{align} 
$M'_\alpha(\bm{k})=\partial_{k_\alpha} M(\bm{k})$, $M''_{\alpha\beta}(\bm{k})=\partial_{k_\alpha}\partial_{k_\beta} M(\bm{k})$,
\begin{align}
\label{eqn17}
W_{\alpha\beta}(\bm{k})=&\frac{3}{8}M''_{\alpha\beta}(\bm{k})-P_{\alpha\beta}(\bm{k})\nonumber\\
=&\frac{3}{8}M''_{\alpha\beta}(\bm{k})-\langle\langle\partial_{k_\alpha} (u_{\bm{k}}^*u_{\bm{k}})|\partial_{k_\beta} (u_{\bm{k}}^*u_{\bm{k}})\rangle\rangle,\nonumber\\
\widetilde{S}_{\alpha\beta}(\bm{k})=&\mathrm{Im}[\langle\langle u_{\bm{k}}u_{\bm{k}}|u_{\bm{k}}\partial_{k_\alpha}\partial_{k_\beta} u_{\bm{k}}\rangle\rangle\nonumber\\
&-\langle\langle u_{\bm{k}}u_{\bm{k}}|(\partial_{k_\alpha} u_{\bm{k}})(\partial_{k_\beta} u_{\bm{k}})\rangle\rangle].
\end{align}
By choosing the same gauge Eq.~(\ref{eqn8}) as in Eq.~(\ref{eqn6}), an effective interaction $H_I=H_{I0}+H_{I1}+H_{I2}$ is also ``gauge-independent" (see Appendix \ref{AppendixA2}). $M(\bm{k})$ itself is gauge-indepedent, while $\widetilde{S}_{\alpha\beta}(\bm{0})$ equals a gauge-independent quantity $S_{\alpha\beta}(\bm{0})$ in this gauge,
\begin{align}
\label{eqn18}
&S_{\alpha\beta}(\bm{k})=\mathrm{Im}[\langle\langle u_{\bm{k}}u_{\bm{k}}|u_{\bm{k}}\partial_{k_\alpha}\partial_{k_\beta} u_{\bm{k}}\rangle\rangle\nonumber\\
&\!\ \!\ -\langle\langle u_{\bm{k}}u_{\bm{k}}|(\partial_{k_\alpha} u_{\bm{k}})(\partial_{k_\beta} u_{\bm{k}})\rangle\rangle-M(\bm{k})\langle u_{\bm{k}}|\partial_{k_\alpha}\partial_{k_\beta} u_{\bm{k}}\rangle].
\end{align}
The argument $\bm{k}$ of the \textit{Bloch wavefunction quantities}, $\Omega_\gamma$, $\mathsf{g}_{\alpha\beta}$, $M$, $M'_\alpha$, $M''_{\alpha\beta}$, $W_{\alpha\beta}$, $S_{\alpha\beta}$, will be omitted when $\bm{k}=\bm{0}$.

\begin{table}[t]
\caption{The permutation and time-reversal symmetries of the gauge-independent second-order tensors characterizing the lattice-induced wavefunction effects. ``+" and ``-" denote even and odd parities, respectively. The parities of $S_{\alpha\beta}$ are different from either $\mathsf{g}_{\alpha\beta}$ or $\Omega_{\alpha\beta}$.}
\label{table1}
\begin{adjustbox}{width=0.35\textwidth}
\footnotesize
\begin{tabular}{c|cc|ccc}
  & $\mathsf{g}_{\alpha\beta}$ & $\Omega_{\alpha\beta}$ & $M''_{\alpha\beta}$ & $W_{\alpha\beta}$ & $S_{\alpha\beta}$ \\
\hline   
Permutation & + & - & + & + & + \\
Time reversal & + & - & + & + & -
\end{tabular}
\end{adjustbox}
\end{table}

Note that the ``double Dirac notation" defined in Eq.~(\ref{eqn11}) is not an inner product between two-body states. Instead, it is a matrix element of the interaction between two-body states. For the $\phi^4$ interaction, the matrix element $M(\bm{k})$ takes the (quartic) inverse participation ratio of the Bloch wavefunction in the unit cell, quantifying the gathering of particles. We can see that the lattice-induced corrections to the interaction are qualitatively distinct from those to the external potential (see Table \ref{table1}). Contrary to $\Omega_{\alpha\beta}$, the three tensors $M''_{\alpha\beta}$, $W_{\alpha\beta}$, $S_{\alpha\beta}$ appearing in the effective interaction are all symmetric tensors. Besides, while $S_{\alpha\beta}$ is symmetric, it is odd under time reversal. So the quantum geometric tensor $\mathcal{B}_{\alpha\beta}=\mathsf{g}_{\alpha\beta}+i\Omega_{\alpha\beta}$ is not sufficient to describe lattice-induced wavefunction effects in the correlated system.

\section{\label{Sec3}Hydrodynamic theory}

In a hydrodynamic theory neglecting quantum corrections, classical equations of motion of a superfluid are determined by $\delta \mathcal{S}=\delta\int dtd^d\bm{r}\mathcal{L}(n,\theta)=0$,
\begin{align}
\label{eqn19}
\partial_t\theta=&-\frac{1}{2m}(\bm{\nabla}\theta)^2+\frac{\nabla^2 n}{4mn}+\frac{(\bm{\nabla} n)^2}{8mn^2}\nonumber\\
&-(U+\frac{1}{2}\mathsf{g}_{\alpha\beta}\partial_\alpha\partial_\beta U)-\frac{1}{2}\bm{\nabla}\theta\cdot(\bm{\nabla}U\times\bm{\Omega})\nonumber\\
&+\mu-gMn-gM'_\alpha n\partial_\alpha\theta-\frac{g}{2}M''_{\alpha\beta}n(\partial_\alpha\theta)(\partial_\beta\theta)\nonumber\\
&-gS_{\alpha\beta}n\partial_\alpha\partial_\beta\theta+gW_{\alpha\beta}\partial_\alpha\partial_\beta n,
\end{align}
\begin{align}
\label{eqn20}
\partial_t n=&-\bm{\nabla}\cdot(\frac{n}{m}\bm{\nabla}\theta+\frac{n}{2}\bm{\nabla}U\times\bm{\Omega})\nonumber\\
&-\partial_\alpha(\frac{g}{2}M'_\alpha n^2+\frac{g}{2}M''_{\alpha\beta}n^2\partial_\beta\theta-gS_{\alpha\beta}n\partial_\beta n).
\end{align}
The chemical potential $\mu$ is included on the right-hand side of Eq.~(\ref{eqn19}). Thereby, we absorb an unobservable uniform phase winding $\partial_t\theta=-\mu$ for ground states without external potential~\cite{Zhai2021}. Eq.~(\ref{eqn20}) is a continuity equation,
\begin{align}
\label{eqn21}
\partial_t n+\bm{\nabla}\cdot(n\bm{v})=0,
\end{align}
where
\begin{align}
\label{eqn22}
v_\alpha=&\frac{1}{m}\partial_\alpha\theta+\frac{1}{2}\epsilon_{\alpha\beta\gamma}(\partial_\beta U)\Omega_\gamma\nonumber\\
&+\frac{g}{2}M'_\alpha n+\frac{g}{2}M''_{\alpha\beta}n\partial_\beta\theta-gS_{\alpha\beta}\partial_\beta n.
\end{align}

\subsection{\label{Sec31}Ground state and collective modes without external potential}

We first simply discuss the ground state of the system when $U(\bm{r})=0$. Suppose $\nabla n$ and $\nabla \theta$ are small. Then at the leading order, Eq.~(\ref{eqn19}) gives $n(\bm{r})=n_0=\frac{\mu}{gM}$. However, $\theta(\bm{r})$ is determined by
\begin{align}
\label{eqn23}
(\frac{1}{m}\delta_{\alpha\beta}+\frac{gn_0}{2}M''_{\alpha\beta})\partial_\beta\theta=-\frac{gn_0}{2}M'_\alpha.
\end{align}
When $M'_\alpha$ is small but finite, $\theta(\bm{r})$ acquires a finite momentum. For a general $M'_\alpha$, we cannot apply the expansion of $\nabla\theta$. In the remainder of the paper, we only consider the case of $M'_\alpha=0$. It can be realized when there is an inversion symmetry, or when there is a $\mathbb{Z}_n$ ($n\geq 3$) rotational symmetry for 2D, or when a 3D lattice belongs to the cubic crystal system. In this case, there is a uniform ground state,
\begin{align}
\label{eqn24}
n(\bm{r})=n_0=\frac{\mu}{gM},\quad \theta(\bm{r})=\theta_0.
\end{align} 

The system has a typical energy scale $\mu$ and a typical momentum scale $\sqrt{m\mu}$, and they can form other typical scales with different dimensions. In the remainder of the paper, for ground states and low-energy collective modes with $U(\bm{r})$, we further assume that reduced by the two typical scales, $\nabla^n U$ and $\nabla^n\theta$ (including $\partial_t^n\theta$) are $n$th-order small quantities (which can be verified \textit{a posteriori}), and the Bloch wavefunction quantities are at most $\mathcal{O}(1)$. Note that $\nabla (n-n_0)$ and $\nabla\theta$ do not need to be in the same order (see Eqs.~(\ref{eqn25}, \ref{eqn31})). Results of the remainder of the paper will be derived in the leading order of this gradient expansion. A correction of the density in Eq.~(\ref{eqn9}) is second-order, so we take $n_{\mathrm{ph}}(\bm{r},t)=n(\bm{r},t)$.

With $n=n_0+\delta n$, we get the leading order of the equations of motion,
\begin{align}
\label{eqn25}
\partial_t^2\theta=&-gM\partial_t\delta n\nonumber\\
=&\frac{gMn_0}{m}\nabla^2\theta+\frac{g^2Mn_0^2}{2}M''_{\alpha\beta}\partial_\alpha\partial_\beta\theta.
\end{align}
The $d\times d$ real symmetric matrix $M''_{\alpha\beta}$ has $d$ eigenvalues $\lambda_1,...,\lambda_d$, so the system is anisotropic in general and we get $d$ sound velocities~\cite{Subasi2022, Lukin2023},
\begin{align}
\label{eqn26}
c_i=\sqrt{\frac{gMn_0}{m}+\frac{g^2Mn_0^2}{2}\lambda_i}=\sqrt{\frac{\mu}{m}+\frac{\lambda_i\mu^2}{2M}}\quad (i=1,...,d).
\end{align}

\subsection{\label{Sec32}Ground state with external potential}

When $U(\bm{r})\neq 0$, a static solution of the system is determined by $\partial_t\theta=\partial_t n=0$. Suppose the ground state of $U(\bm{r})\neq 0$ is static and adiabatically connected to the ground state of $U(\bm{r})=0$. Denote the ground state as $b(\bm{r})=\sqrt{n_0(\bm{r})}e^{i\theta_0(\bm{r})}$. From Eq.~(\ref{eqn19}), in the leading order we get
\begin{align}
\label{eqn27}
n_0(\bm{r})=\left\{\begin{array}{cc}\frac{\mu-U(\bm{r})}{gM}&\mathrm{when}\quad \mu>U(\bm{r})\\ 0&\mathrm{when}\quad\mu\leq U(\bm{r})\end{array}\right..
\end{align}
When $n(\bm{r})$ is fixed, the Lagrangian $\mathcal{L}$ becomes a quadratic form of $\nabla\theta$. To minimize the quadratic form, we cannot simply require $\bm{v}(n_0,\theta_0)=\bm{0}$, because $n\bm{v}$ is not rotation-free in general. In the leading order, to solve $\theta_0(\bm{r})$, we can take $n$ in Eq.~(\ref{eqn20}) to be $n_0$. As $\bm{\nabla} U$ is proportional to $\bm{\nabla} n_0$, the antisymmetric term of Berry curvature vanishes. Then $\theta_0(\bm{r})$ is determined by
\begin{align}
\label{eqn28}
&[\frac{n_0}{m}\delta_{\alpha\beta}+\frac{g}{2}M''_{\alpha\beta}n_0^2]\partial_\alpha\partial_\beta\theta_0+[\frac{1}{m}\delta_{\alpha\beta}\nonumber\\
&\!\ \!\ +gM''_{\alpha\beta}n_0](\partial_\alpha n_0)(\partial_\beta\theta_0)=\frac{1}{2}gS_{\alpha\beta}\partial_\alpha\partial_\beta n_0^2.
\end{align}

When the time-reversal symmetry is broken such that $S_{\alpha\beta}\neq 0$, it is nontrivial to determine $\theta_0(\bm{r})$. In the remainder of the paper, we focus on a simple case, where the external potential is central, i.e.~$U(\bm{r})=U(r)$. Furthermore, suppose the lattice belongs to the cubic crystal system in 3D or there is a $\mathbb{Z}_n$ ($n\geq 3$) rotational symmetry in 2D. For those highly symmetric cases, $M'_\alpha=0$, $\mathsf{g}_{\alpha\beta}=\mathsf{g}\delta_{\alpha,\beta}$, $M''_{\alpha\beta}=M''\delta_{\alpha,\beta}$, $W_{\alpha\beta}=W\delta_{\alpha,\beta}$, $S_{\alpha\beta}=S\delta_{\alpha,\beta}$; $\bm{\Omega}=\Omega\hat{\bm{e}}_z$ for 2D and $\bm{\Omega}=\bm{0}$ for 3D. The stability of the ground state without $U(r)$ requires
\begin{align}
\label{eqn29}
\eta\equiv\frac{M''m\mu}{M}>-2. 
\end{align}
Suppose the ground-state configuration is isotropic. Then we get the rotation-free part of the velocity for $n(\bm{r})=n(r)$, $\theta(\bm{r})=\theta(r)$,
\begin{align}
\label{eqn30}
\tilde{\bm{v}}&\equiv\bm{v}-\frac{1}{2}(\bm{\nabla}U)\times\bm{\Omega}\nonumber\\
&=(\frac{1}{m}+\frac{gM''}{2}n)\bm{\nabla}\theta-gS\bm{\nabla} n.
\end{align}
The ground state is given by $\tilde{\bm{v}}(n_0,\theta_0)=\bm{0}$, as a term $n\bm{\nabla}\theta\cdot(\bm{\nabla}U\times\bm{\Omega})$ in the Lagrangian $\mathcal{L}$ vanishes. For $\mu>U(r)$, we get
\begin{align}
\label{eqn31}
[\frac{1}{m}+\frac{M''}{2M}(\mu-U)]\bm{\nabla}\theta_0=-\frac{S}{M}\bm{\nabla}U.
\end{align}
According to Eqs.~(\ref{eqn27}, \ref{eqn31}), small $\nabla U$ justifies the momentum expansion of $\nabla n_0$ and $\nabla\theta_0$.

Eq.~(\ref{eqn27}) is the same as the Thomas-Fermi distribution~\cite{Zhai2021, Li2017} (except a constant factor $M$). However, Eq.~(\ref{eqn31}) gives a correction to the Thomas-Fermi distribution, which cannot be obtained when the interaction between atoms is treated as a mean field~\cite{Li2017, Price2013}. $\nabla\theta_0$ becomes nonzero and in the same order as $\nabla U$. Although an equilibrium distribution of $\theta_0(\bm{r})$ may not be observed directly, there are observable effects when the equilibrium distribution depends on time due to manipulation of the optical lattice. An observable dynamic process thereby will be discussed in Sec.~\ref{Sec422}.

\subsection{\label{Sec33}Collective modes with external potential}

Consider collective modes near the isotropic static ground state, $n(\bm{r},t)=n_0(r)+\delta n(\bm{r},t)$, $\theta(\bm{r},t)=\theta_0(r)+\delta \theta(\bm{r},t)$. We take the double expansion of derivatives of $U$ and $\delta\theta$ in Eqs.~(\ref{eqn19}, \ref{eqn20}), and calculate leading-order equations of motion. Similar to the case of $U(r)=0$, $\delta n$ is in the same order as $\partial_t\delta\theta$,
\begin{align}
\label{eqn32}
\partial_t\delta\theta=-gM\delta n,
\end{align}
where we have used the condition that $\partial_t\delta\theta=0$ when $\delta\theta=\delta n=0$. For $\mu>U(r)$, Eq.~(\ref{eqn21}) gives
\begin{align}
\label{eqn33}
\partial_t^2\delta n=&-\bm{\nabla}\cdot(n_0\partial_t\bm{v})-\bm{\nabla}\cdot(\delta n\partial_t\bm{v})\nonumber\\
&-\bm{\nabla}\cdot[(\partial_t\delta n)\frac{1}{2}(\bm{\nabla} U)\times\bm{\Omega}]-\bm{\nabla}\cdot[(\partial_t\delta n)\tilde{\bm{v}}]\nonumber\\
\approx&-\bm{\nabla}\cdot(n_0\partial_t\bm{v}).
\end{align}
Here the first term on the right-hand side is in the same order as $\partial_t^2\delta n$, which will be verified in Eq.~(\ref{eqn34}). The second term is neglected because $\delta n$ makes it one order higher than $\partial_t^2\delta n$. The third term is neglect because $\bm{\nabla} U$ makes it one order higher than $\partial_t^2\delta n$, although a correction of the Berry curvature was discussed in previous works~\cite{Li2017, Price2013}. The last term is neglected because $\tilde{\bm{v}}(n_0,\theta_0)=0$, so terms in $\tilde{\bm{v}}$ are proportional to $\nabla\delta\theta$ or $\delta n$, which makes the last term one order higher than $\partial_t^2\delta n$. Then for $\mu>U(r)$, $\delta n(\bm{r},t)$ is given by
\begin{align}
\label{eqn34}
\partial_t^2\delta n\approx&-\bm{\nabla}\cdot[\frac{\mu-U}{gM}(\frac{1}{m}\bm{\nabla}\partial_t\delta\theta+\frac{g}{2}M''\frac{\mu-U}{gM}\bm{\nabla}\partial_t\delta\theta)]\nonumber\\
=&\bm{\nabla}\cdot(\mu-U)[\frac{1}{m}\bm{\nabla}\delta n+\frac{M''}{2M}(\mu-U)\bm{\nabla}\delta n]\nonumber\\
=&\frac{\mu-U}{m}[1+\frac{\eta}{2}(1-\frac{U}{\mu})]\nabla^2\delta n\nonumber\\
&-\frac{\bm{\nabla} U}{m}\cdot[1+\eta(1-\frac{U}{\mu})]\bm{\nabla}\delta n.
\end{align}
Here we also neglected the terms proportional to $\bm{\nabla}\partial_t\delta n$ and $(\partial_t\delta n)\bm{\nabla}\delta\theta$ in $\partial_t\bm{v}$, because their orders are one higher than $\bm{\nabla}\partial_t\delta\theta$. Compared with Eqs.~(\ref{eqn26}, \ref{eqn27}), we know that the first term in Eq.~(\ref{eqn34}) is to replace the constant density $n_0$ by the density distribution $n_0(\bm{r})$. The two terms in Eq.~(\ref{eqn34}) are both important when the momentum scales of $U$ and $\delta n$ are in the same order.

\subsection{\label{Sec34}Discussion on quantum fluctuations}
In this section, we derive the lattice-induced corrections to the hydrodynamic theory in the classical-field limit, i.e.~$\delta\mathcal{S}=0$. Besides, quantum fluctuations may also give corrections. Typically, the quantum-fluctuation corrections are small (e.g.~about 1\%)~\cite{Pitaevskii1998, Braaten1999}. However, it should be clarified in what limit our treatment is justified. The ratio between the quantum correction to the particle density $n_{\mathrm{q}}$ and classical density $n_{\mathrm{cl}}$ vanish in a weakly interacting limit ($d=2,3$),
\begin{align}
\label{eqn34a}
\frac{n_{\mathrm{q}}}{n_{\mathrm{cl}}}\sim\frac{\xi_h^{-d}}{\mu/g}\sim\frac{(m\mu)^{\frac{d}{2}}}{\mu/g}=m^{\frac{d}{2}}\mu^{\frac{d}{2}-1}g\rightarrow 0,
\end{align}
where the particle density $n=n_{\mathrm{cl}}+n_{\mathrm{q}}$, and $\xi_h$ is the typical length scale (i.e.~healing length~\cite{Zhai2021}) of the superfluid state. Eq.~(\ref{eqn34a}) guarantees the applicability of the hydrodynamic theory. On the other hand, the lattice-induced corrections may survive in this limit. Suppose the interaction strength $g$ decreases while the particle density $n$ increases, such that the chemical potential $\mu$ keeps in the same order. Then from the classical equations of motion Eqs.~(\ref{eqn19}, \ref{eqn20}), or from the results Eqs.~(\ref{eqn26}, \ref{eqn27}, \ref{eqn31}, \ref{eqn34}), we can see that the lattice-induced corrections remain in the same order, despite the suppression of the quantum corrections. Therefore, it is fair to stay on this limit and neglect the quantum fluctuations in this paper. Our derivations can be generalized when further taking the quantum fluctuations into consideration.

\begin{figure*}[t]
\centering
\subfigure[ ]{
\begin{minipage}{0.23\textwidth}
\centering
\includegraphics[width=\textwidth]{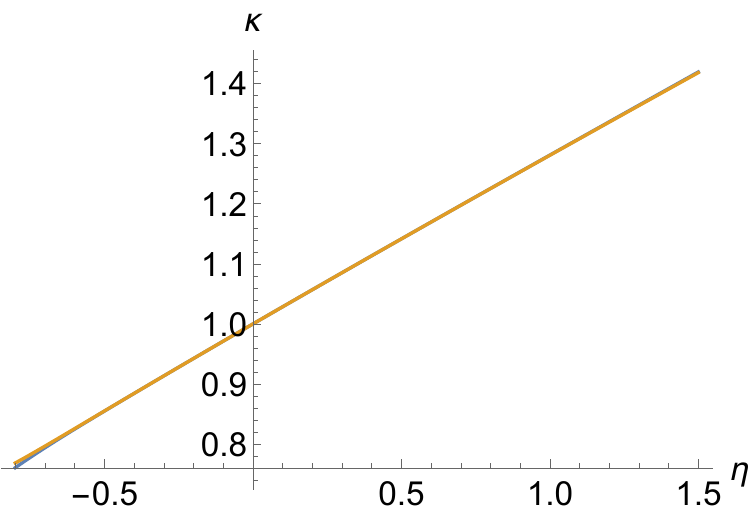}
\end{minipage}
}
\subfigure[ ]{
\begin{minipage}{0.23\textwidth}
\centering
\includegraphics[width=\textwidth]{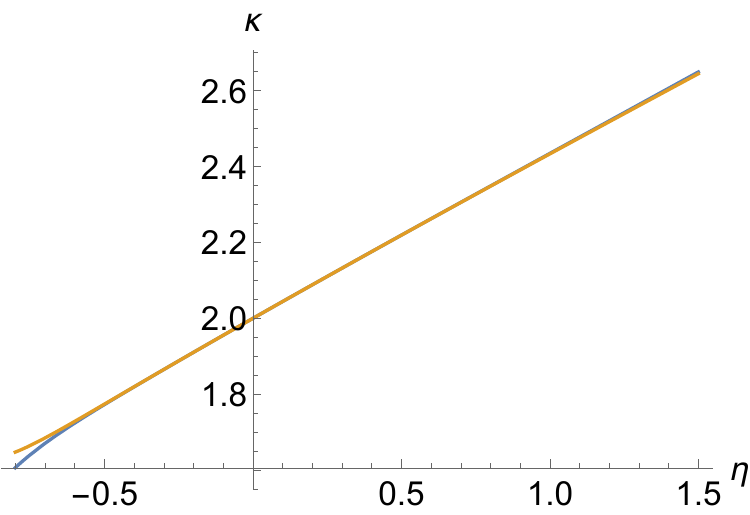}
\end{minipage}
}
\subfigure[ ]{
\begin{minipage}{0.23\textwidth}
\centering
\includegraphics[width=\textwidth]{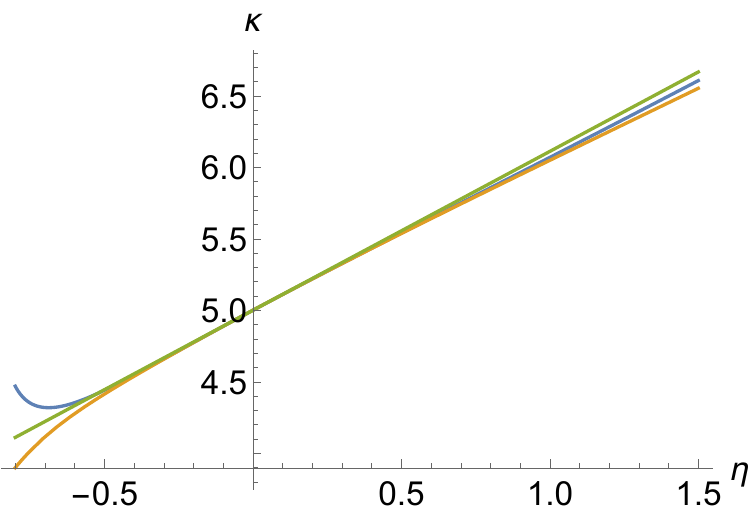}
\end{minipage}
}
\subfigure[ ]{
\begin{minipage}{0.23\textwidth}
\centering
\includegraphics[width=\textwidth]{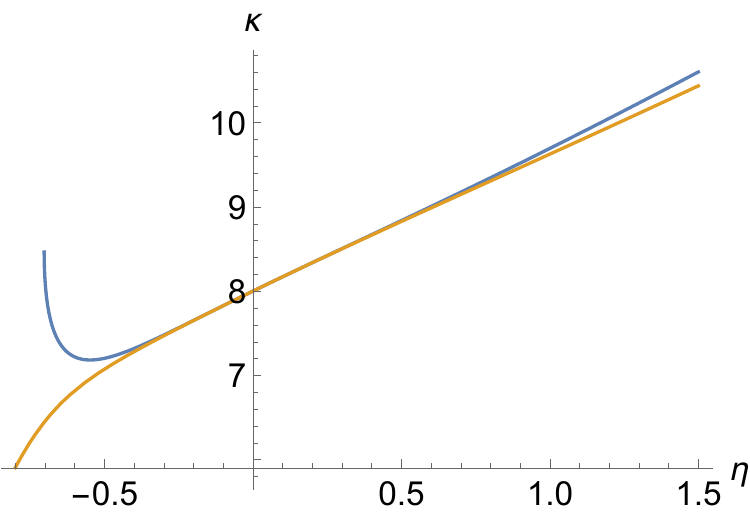}
\end{minipage}
}

\subfigure[ ]{
\begin{minipage}{0.23\textwidth}
\centering
\includegraphics[width=\textwidth]{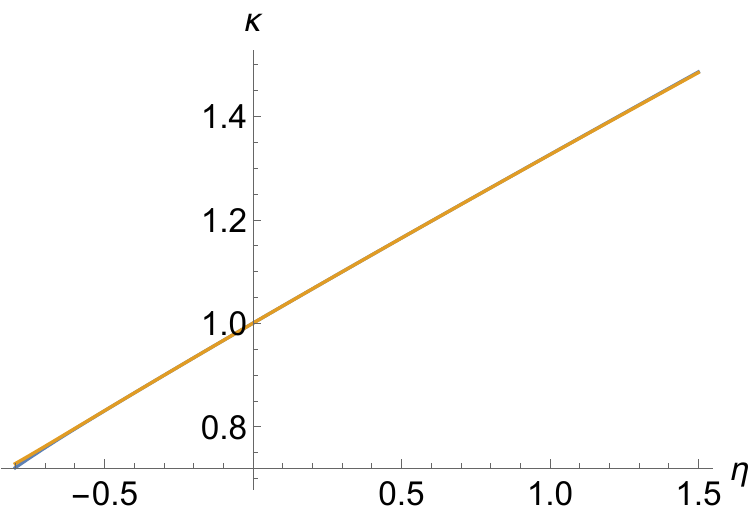}
\end{minipage}
}
\subfigure[ ]{
\begin{minipage}{0.23\textwidth}
\centering
\includegraphics[width=\textwidth]{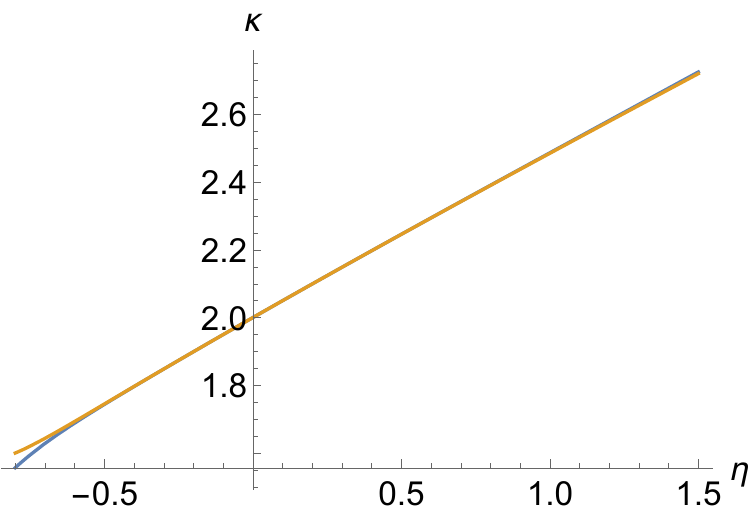}
\end{minipage}
}
\subfigure[ ]{
\begin{minipage}{0.23\textwidth}
\centering
\includegraphics[width=\textwidth]{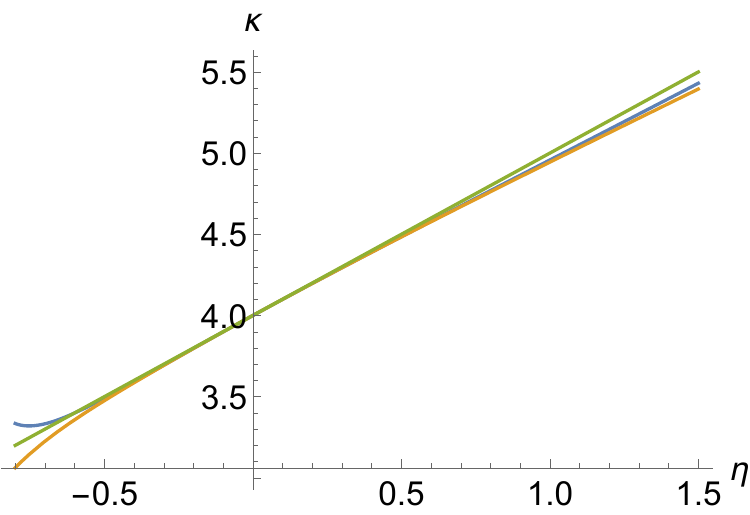}
\end{minipage}
}
\subfigure[ ]{
\begin{minipage}{0.23\textwidth}
\centering
\includegraphics[width=\textwidth]{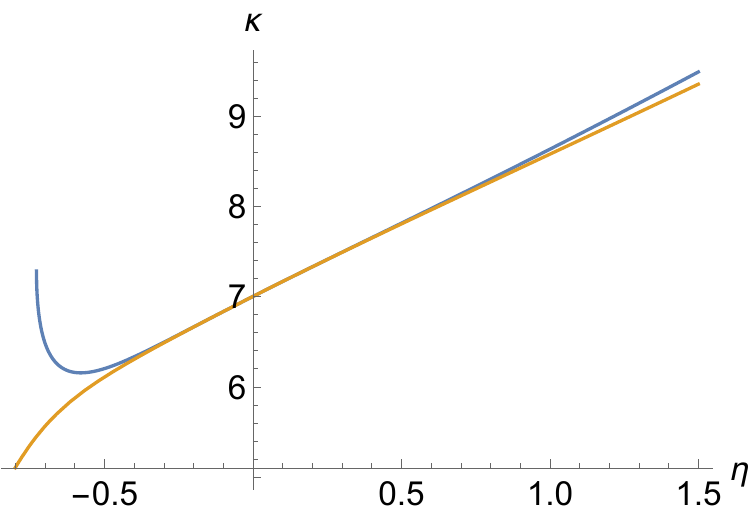}
\end{minipage}
}
\caption{\label{fig1} Numerical results of eigenfrequencies of collective modes ($\omega_{\mathsf{n},\mathsf{j}}\equiv\omega_0\sqrt{\kappa}$) in an isotropic harmonic trap as functions of the quantity $\eta=\frac{M''m\mu}{M}$. Figs.~(a-d) are for 3D ($\mathsf{j}=\ell$) and Fig.~(e-h) are for 2D ($\mathsf{j}=|\mathsf{m}|$). Parameters are taken to be $(\mathsf{n},\mathsf{j})=(0,1)$ in Figs.~(a,e), $(\mathsf{n},\mathsf{j})=(0,2)$ in Figs.~(b,f), $(\mathsf{n},\mathsf{j})=(1,0)$ in Figs.~(c,g), and $(\mathsf{n},\mathsf{j})=(1,1)$ in Figs.~(d,h). Blue lines are for $\mathsf{k}_0=4$ and yellow lines are for $\mathsf{k}_0=5$ defined in Eq.~(\ref{eqn41}). The results are convergent where the differences between the blue line and the yellow line are much smaller than the averages. The two lines almost coincide for most of the data of $\eta>-1$, thus it shows good numerical convergence. When $\mathsf{n}$ or $|\eta|$ increases, larger $\mathsf{k}_0$ is needed for convergence. Besides, we plot green lines in Fig.~(c,g) according to the first-order perturbation theory (see Eq.~(\ref{eqn48}) with $\eta\approx m\mu M''$), where $\kappa$ is linear in $\eta$. The perturbative results are consistent with the numerical results.}
\end{figure*}

\section{\label{Sec4}Superfluid dynamics in a harmonic trap}

In cold-atom systems, $U(r)$ is commonly taken as an isotropic harmonic potential~\cite{Zhai2021},
\begin{align}
\label{eqn35}
U(r)=\frac{1}{2}m\omega_0^2 r^2\equiv \mu\frac{r^2}{R^2}.
\end{align}
Here $\omega_0\ll\mu\lesssim J$ and $k_0\equiv\frac{1}{R}\ll\sqrt{m\mu}\lesssim k_M$ such that the gradient expansion (i.e.~momentum and frequency expansion) is applicable for $r<R$. $J$ is the bandwidth of the lowest band and $k_M$ is the momentum range of the first Broullion zone.

\subsection{\label{Sec41}Collective modes}

Define $\rho(\frac{\bm{r}}{R},t)\equiv\delta n(\bm{r},t)$. Solutions of Eqs.~(\ref{eqn34}, \ref{eqn35}) depend on $\eta$. For the 3D case with spherical coordinates $\bm{r}=(r,\theta,\phi)$,
\begin{align}
\label{eqn36}
\rho(\bm{r},t)=\sum_{\mathsf{n},\ell,\mathsf{m}}\sum_{\xi=\pm}e^{-i\xi\omega_{\mathsf{n},\ell} t}\mathsf{a}_{\mathsf{n},\ell,\mathsf{m},\xi}\rho_{\mathsf{n},\ell}(r)Y_{\ell,\mathsf{m}}(\theta,\phi),
\end{align}
where $Y_{\ell,m}$ are spherical harmonics. Then we get a one-dimensional eigenequation for $\rho_{\mathsf{n},\ell}(r)$ with $0<r<1$,
\begin{align}
\label{eqn37}
&-\omega_{\mathsf{n},\ell}^2\rho_{\mathsf{n},\ell}=\frac{1}{2}\omega_0^2(1-r^2)[1+\frac{\eta}{2}(1-r^2)][\frac{d^2}{d r^2}+\frac{2}{r}\frac{d}{d r}\nonumber\\
&\!\ \!\ \!\ \!\ -\frac{\ell(\ell+1)}{r^2}]\rho_{\mathsf{n},\ell}-\omega^2_0 r[1+\eta(1-r^2)]\frac{d}{d r}\rho_{\mathsf{n},\ell}.
\end{align}
For the 2D case with polar coordinates $\bm{r}=(r,\phi)$,
\begin{align}
\label{eqn38}
\rho(\bm{r},t)=\sum_{\mathsf{n},\mathsf{m}}\sum_{\xi=\pm}e^{-i\xi\omega_{\mathsf{n},|\mathsf{m}|} t}\mathsf{a}_{\mathsf{n},\mathsf{m},\xi}\rho_{\mathsf{n},|\mathsf{m}|}(r)e^{i\mathsf{m}\phi}.
\end{align}
Then we get
\begin{align}
\label{eqn39}
&-\omega_{\mathsf{n},|\mathsf{m}|}^2\rho_{\mathsf{n},|\mathsf{m}|}=\frac{1}{2}\omega_0^2(1-r^2)[1+\frac{\eta}{2}(1-r^2)][\frac{d^2}{d r^2}+\frac{1}{r}\frac{d}{d r}\nonumber\\
&\!\ \!\ \!\ \!\ -\frac{m^2}{r^2}]\rho_{\mathsf{n},|\mathsf{m}|}-\omega^2_0 r[1+\eta(1-r^2)]\frac{d}{d r}\rho_{\mathsf{n},|\mathsf{m}|}.
\end{align}
In Eqs.~(\ref{eqn36}, \ref{eqn38}), $\mathsf{n}$ takes nonnegative integers.

When $\eta\neq 0$, eigenfrequencies (eigenenergies) $\omega_{\mathsf{n},\mathsf{j}}$ cannot be obtained analytically, where $\mathsf{j}\equiv\ell$ for $d=3$ and $\mathsf{j}\equiv|\mathsf{m}|$ for $d=2$. Nevertheless, when $\eta>-1$, solutions take the form of an infinite series, 
\begin{align}
\label{eqn40}
\rho_{\mathsf{n},\mathsf{j}}(r)=\sum_{\mathsf{k}=0}^\infty\alpha_{\mathsf{n},\mathsf{j},2\mathsf{k}}r^{2\mathsf{k}+\mathsf{j}},
\end{align}
and the eigenfrequencies are solvable numerically~\cite{SM2}. According to an asymptotic analysis of $\alpha_{\mathsf{n},\mathsf{j},2\mathsf{k}}$ (see Appendix \ref{AppendixB2}), we can iteratively determine $\frac{\alpha_{\mathsf{n},\mathsf{j},2\mathsf{k}+2}}{\alpha_{\mathsf{n},\mathsf{j},2\mathsf{k}}}$ as functions of $\omega_{\mathsf{n},\mathsf{j}}$ for $\mathsf{k}\leq\mathsf{k}_0$, and obtain $\omega_{\mathsf{n},\mathsf{j}}$ (see Fig.~\ref{fig1}) by requiring
\begin{align}
\label{eqn41}
\frac{\alpha_{\mathsf{n},\mathsf{j},2\mathsf{k}_0+2}}{\alpha_{\mathsf{n},\mathsf{j},2\mathsf{k}_0}}=\frac{\mathsf{k}_0}{\mathsf{k}_0+1}\frac{\eta}{\eta+2},
\end{align}
with a certain integer $\mathsf{k}_0$ large enough for convergence. For several low-energy levels, i.e. $\mathsf{n}$ and $\mathsf{j}$ are not large, the eigenfrequency $\omega_{\mathsf{n},\mathsf{j}}$ is in the same order as $\omega_0$. According to Eqs.~(\ref{eqn32}, \ref{eqn34}) with $\omega_0\ll\mu$, this justifies the momentum and frequency expansion of $\delta n_0$ and $\delta\theta_0$ for the low-energy collective modes. When $\eta=0$, Eq.~(\ref{eqn41}) returns to the condition that the series becomes finite~\cite{Stringari1996}, and the numerical results in Fig.~\ref{fig1} are consistent with those without the lattice.

Besides, collective modes are normalizable and satisfy an orthonormal relation~\cite{Mathews1971} (see Appendix \ref{AppendixB3} for a proof), 
\begin{align}
\label{eqn42}
\int_0^1 r^{d-1}\rho_{\mathsf{n}_1,\mathsf{j}}^*(r)\rho_{\mathsf{n}_2,\mathsf{j}}(r)dr=\delta_{\mathsf{n}_1,\mathsf{n}_2}.
\end{align}
The orthonormal condition is independent of $\eta$, which enables perturbative solutions of the collective modes in Sec.~\ref{Sec421}.

\subsection{\label{Sec42}Breathing mode with weak lattice-induced wavefunction effects}

From Eqs.~(\ref{eqn27}, \ref{eqn29}, \ref{eqn31}) and Fig.~\ref{fig1}, in the leading order of the gradient expansion, we know $M_d\equiv M-1$, $M''$, and $S$ are already involved in the dynamics of the superfluid in the isotropic harmonic trap~\cite{SM3}. To demonstrate more explicit relations between the three Bloch wavefunction quantities and experimental observables, suppose the lattice-induced wavefunction effects are weak,
\begin{align}
\label{eqn43}
M_d\equiv M-1\ll 1,\quad m\mu|M''|\ll 1,\quad m\mu|S|\ll 1.
\end{align}
The last condition in Eq.~(\ref{eqn43}) can be weakened and substituted by $m\omega_0|S|\ll 1$, which has been automatically satisfied by previously imposed conditions $m\mu|S|=\mathcal{O}(1)$ and $\omega_0\ll\mu$. We will keep the leading order of the three quantities, which leads to $|\eta|\approx|M''|m\mu\ll 1$.

The weak wavefunction effects can be realized by a weak lattice potential. They can also be effectively realized in a tight-binding limit by weakly breaking sublattice symmetry, where the integrals of $\bm{r}$ in the definition Eq.~(\ref{eqn11}) should be renormalized and become summations of a sublattice index (see Appendix \ref{AppendixC1}). Here we define the sublattice symmetry physically, i.e.~by permuting the sublattices, which may be different from a conventional one~\cite{Schnyder2008, Schnyder2009, Kitaev2009, Ryu2010}. 

As an example, we focus on the lattice-induced wavefunction effects on the breathing mode ($\mathsf{n}=1$, $\mathsf{j}=0$)~\cite{Zhai2021}. As explained below Eq.~(\ref{eqn41}), the momentum and frequency expansion is valid for the breathing mode. Without the wavefunction effects, (zeroth-order) eigenvalues ($\kappa_{\mathsf{n},\mathsf{j}}\equiv\frac{\omega^2_{\mathsf{n},\mathsf{j}}}{\omega_0^2}$) and normalized eigenfunctions ($\rho_{\mathsf{n},\mathsf{j}}$) of Eqs.~(\ref{eqn37}, \ref{eqn39}) for the breathing mode are given by (See Appendix \ref{AppendixB1} and Ref.~\cite{Stringari1996})
\begin{align}
\label{eqn44}
\kappa^{(0)}_{1,0}=\left\{\begin{array}{cc}5&\quad\mathrm{for}\quad\mathrm{3D}\\ 4&\quad\mathrm{for}\quad\mathrm{2D}\end{array}\right.,
\end{align}
\begin{align}
\label{eqn45}
\rho^{(0)}_{1,0}=\left\{\begin{array}{cc}\frac{3\sqrt{7}}{2}(1-\frac{5}{3}r^2) &\quad\mathrm{for}\quad\mathrm{3D}\\ \sqrt{6}(1-2r^2) &\quad\mathrm{for}\quad\mathrm{2D}\end{array}\right..
\end{align}

\subsubsection{\label{Sec421}Eigenfrequency}

At the first order, $\kappa_{1,0}$ should be a linear function of $\eta$, i.e. a linear function of $M''$. We can apply a perturbation theory similar to that in quantum mechanics and get the first-order correction to the eigenvalue $\kappa_{1,0}$,
\begin{align}
\label{eqn46}
\kappa^{(1)}_{1,0}=\int_0^1 r^{d-1}\rho^{(0)}_1\hat{L}^{(1)}\rho^{(0)}_1 dr=\left\{\begin{array}{cc}\frac{10}{9}\eta&\quad\mathrm{for}\quad\mathrm{3D}\\ \eta&\quad\mathrm{for}\quad\mathrm{2D}\end{array}\right.,
\end{align}
where the first-order operator $\hat{L}^{(1)}$ with $\ell=0$ is given by
\begin{align}
\label{eqn47}
\hat{L}^{(1)}=\left\{\begin{array}{cc}\begin{array}{l}-\frac{\eta}{4}(1-r^2)^2\frac{d^2}{d r^2} \\ \!\ \!\ \!\ \!\ -\frac{\eta}{2}(1-r^2)(1-3r^2)\frac{1}{r}\frac{d}{dr}\end{array} &\quad\mathrm{for}\quad\mathrm{3D}\\ \begin{array}{l}-\frac{\eta}{4}(1-r^2)^2\frac{d^2}{d r^2} \\ \!\ \!\ \!\ \!\ -\frac{\eta}{4}(1-r^2)(1-5r^2)\frac{1}{r}\frac{d}{dr}\end{array}&\quad\mathrm{for}\quad\mathrm{2D}\end{array}\right..
\end{align}
So the first-order eigenfrequency of the breathing mode reads
\begin{align}
\label{eqn48}
\omega_{1,0}=\left\{\begin{array}{cc}\sqrt{5}\omega_0(1+\frac{1}{9}m\mu M'')&\quad\mathrm{for}\quad\mathrm{3D}\\ 2\omega_0(1+\frac{1}{8}m\mu M'')&\quad\mathrm{for}\quad\mathrm{2D}\end{array}\right..
\end{align}

\subsubsection{\label{Sec422}Amplitude and initial phase in a dynamic process}

To further show the effects of $M_d$ and $S$, we consider a dynamic process. Atoms are initially at the ground state with the harmonic potential, satisfying the Thomas-Fermi distribution. There is no lattice or a tight-binding lattice with negligible Bloch wavefunction effects, i.e.~$M_d=M''=S=0$. At time $t=0$, we suddenly open or change the lattice potential to make the weak lattice-induced wavefunction effects. Then the ground-state distribution is substituted by Eqs.~(\ref{eqn27}, \ref{eqn31}), and the atoms are at an excited state.

Suppose for $t<0$, the chemical potential, the effective mass of atoms, the typical frequency of the harmonic traps, and the traping radius of the atoms are $\tilde{\mu}$, $\tilde{m}$, $\tilde{\omega}$, $\tilde{R}$. For $t>0$, they become $\mu$, $m$, $\omega$, $R$. Because $U(r)$ does not change, we have
\begin{align}
\label{eqn49}
\frac{1}{2}m\omega_0^2 r^2=\frac{1}{2}\tilde{m}\tilde{\omega}_0^2 r^2,
\end{align}
\begin{align}
\label{eqn50}
\mu=\frac{1}{2}m\omega_0^2 R^2,\quad \tilde{\mu}=\frac{1}{2}\tilde{m}\tilde{\omega}_0^2\tilde{R}^2.
\end{align}
Besides, suppose the amount of atoms remains invariant during the process,
\begin{align}
\label{eqn51}
\frac{\mu}{gM}\int_0^R (1-\frac{r^2}{R^2})r^{d-1}dr=\frac{\tilde{\mu}}{g}\int_0^{\tilde{R}}(1-\frac{r^2}{\tilde{R}^2})r^{d-1}dr.
\end{align}
Combining Eqs.~(\ref{eqn49}--\ref{eqn51}) with Eqs.~(\ref{eqn27}, \ref{eqn31}), we get initial conditions at $t=0$,
\begin{align}
\label{eqn52}
\rho(\bm{r},0)\approx\left\{\begin{array}{cc}\frac{\mu}{g}M_d(\frac{3}{5}-r^2)&\quad\mathrm{for}\quad\mathrm{3D}\\ \frac{\mu}{g}M_d(\frac{1}{2}-r^2)&\quad\mathrm{for}\quad\mathrm{2D}\end{array}\right.,
\end{align}
\begin{align}
\label{eqn53}
\frac{d}{dr}\delta\theta(\bm{r},0)\approx 2m\mu S\frac{r}{R^2},
\end{align}
\begin{align}
\label{eqn54}
\partial_t\delta n(\bm{r},0)\approx-\bm{\nabla}\cdot(n_0\bm{v})\approx-\bm{\nabla}\cdot(\bm{\hat{e}_r}\frac{n_0}{m}\frac{d}{dr}\delta\theta).
\end{align}
Eq.~(\ref{eqn52}) is not valid for $\frac{\tilde{R}}{R}<r<1$. However, that region leads to a higher-order contribution. Eqs.~(\ref{eqn53}, \ref{eqn54}) further gives
\begin{align}
\label{eqn55}
\partial_t\rho(\bm{r},0)=\left\{\begin{array}{cc}-\frac{m\mu}{g}\omega_0^2S(3-5r^2)&\quad\mathrm{for}\quad\mathrm{3D}\\ -\frac{2m\mu}{g}\omega_0^2S(1-2r^2)&\quad\mathrm{for}\quad\mathrm{2D}\end{array}\right..
\end{align}

To satisfy the initial conditions Eqs.~(\ref{eqn52}, \ref{eqn55}), for $t>0$, $\rho(r,t)$ takes a form,
\begin{align}
\label{eqn56}
\rho(\bm{r},t)=\sum_{\mathsf{n}}\mathsf{b}_{\mathsf{n},0}\rho_{\mathsf{n},0}(r)\mathrm{cos}(\omega_{\mathsf{n},0} t+\varphi_{\mathsf{n},0}),
\end{align}
where amplitudes ($\mathsf{b}_{\mathsf{n},0}$) and initial phases ($\varphi_{\mathsf{n},0}$) are determined,
\begin{align}
\label{eqn57}
\mathsf{b}_{\mathsf{n},0}\mathrm{cos}\varphi_{\mathsf{n},0}=\int_0^1 r^{d-1}\rho(r,0)\rho_{\mathsf{n},0}(r)dr,
\end{align}
\begin{align}
\label{eqn58}
\omega_{\mathsf{n},0}\mathsf{b}_{\mathsf{n},0}\mathrm{sin}\varphi_{\mathsf{n},0}=-\int_0^1 r^{d-1}\partial_t\rho(r,0)\rho_{\mathsf{n},0}(r)dr.
\end{align}
At the leading order, we can take $\rho_{\mathsf{n},0}(r)=\rho_{\mathsf{n},0}^{(0)}(r)$. Then we get the amplitude and initial phase of the breathing mode,
\begin{align}
\label{eqn59}
\mathsf{b}_{1,0}\mathrm{cos}\varphi_{1,0}=\left\{\begin{array}{cc}\frac{2\mu M_d}{5\sqrt{7}g}&\quad\mathrm{for}\quad\mathrm{3D}\\ \frac{\mu M_d}{2\sqrt{6}g}&\quad\mathrm{for}\quad\mathrm{2D}\end{array}\right.,
\end{align}
\begin{align}
\label{eqn60}
\mathsf{b}_{1,0}\mathrm{sin}\varphi_{1,0}=\left\{\begin{array}{cc}\frac{2m\mu\omega_0 S}{\sqrt{35}g}&\quad\mathrm{for}\quad\mathrm{3D}\\ \frac{m\mu\omega_0 S}{\sqrt{6}g}&\quad\mathrm{for}\quad\mathrm{2D}\end{array}\right..
\end{align}
The phase shift $\varphi_{1,0}$ can be observed by a corresponding time shift $\tau_{1,0}\equiv\frac{\varphi_{1,0}}{\omega_0}$. A nonzero phase shift $\varphi_{1,0}$ is a characteristic of time-reversal symmetry breaking. 

When $M_d$ and $m\mu S$ are in the same order, despite $|\varphi_{1,0}|\ll 1$, it is possible to observe the time shift as $\tau_{1,0}=\mathcal{O}(\mu^{-1})$. Besides, even if $|\varphi_{1,0}|\ll 1$, when the corrections to $\omega_{1,0}$ and $\rho_{1,0}$ in the next-to-leading order is considered, a correction to $\varphi_{1,0}$ is also in the next-to-leading order.

\section{\label{Sec5}Tight-binding example}

A nonzero $S$ can be generally realized by a synthetic magnetic field breaking the time-reversal symmetry. For a toy model showing essential physics, we consider a 2D tight-binding model of a bilayer square lattice in a nonuniform magnetic field, where the $\mathbb{Z}_4$ rotational symmetry remains intact,
\begin{align}
\label{eqn61}
H_0=&\sum_{\bm{i},\bm{j}}(a^\dagger_{\bm{i},A},a^\dagger_{\bm{i},B})(\bm{h}_{0})_{\bm{i},\bm{j}}\left(\begin{array}{c}a_{\bm{j},A}\\ a_{\bm{j},B}\end{array}\right),\nonumber\\
\frac{(\bm{h}_{0})_{\bm{i},\bm{j}}}{a_L^2}=&\left(\begin{array}{cc}\Delta\delta_{\bm{i},\bm{j}}-t_1\delta_{\langle\bm{i},\bm{j}\rangle} & -t_2\delta_{\bm{i},\bm{j}}-t_3e^{-i\theta_3}\delta_{\langle\bm{i},\bm{j}\rangle} \\ -t_2\delta_{\bm{i},\bm{j}}-t_3e^{i\theta_3}\delta_{\langle\bm{i},\bm{j}\rangle} & -\Delta\delta_{\bm{i},\bm{j}}-t_1\delta_{\langle\bm{i},\bm{j}\rangle}\end{array}\right).
\end{align}
Here $\bm{i}$ and $\bm{j}$ are two-dimensional site positions, $A$ and $B$ are two layers, and $a_L$ is the edge length of a unit cell. $\delta_{\langle\bm{i},\bm{j}\rangle}=1$ when $\bm{i}$ and $\bm{j}$ are the nearest neighbor sites in the 2D plane and vanishes for other cases. $t_{1,2,3}$ are positive hopping coefficients. $\Delta$ is an on-site potential breaking the sublattice symmetry of the two layers. $\theta_3$ is a phase of the complex interlayer hopping between neighboring 2D sites.  When $\mathrm{sin}\theta_3\neq 0$, the complex hopping breaks both the time-reversal symmetry and sublattice symmetry. The complex hopping can be induced by an in-plane synthetic magnetic field rotating around every vertical bond. 

The Hamiltonian in the momentum space reads
\begin{align}
\label{eqn62}
(\bm{H}_0)_{\bm{k}}=&\Delta\bm{\sigma}_z-2t_1[\mathrm{cos}(k_x a_L)+\mathrm{cos}(k_y a_L)]\bm{\sigma}_0-t_2\bm{\sigma}_x\nonumber\\
&-2t_3\mathrm{cos}\theta_3[\mathrm{cos}(k_x a_L)+\mathrm{cos}(k_y a_L)]\bm{\sigma}_x\nonumber\\
&-2t_3\mathrm{sin}\theta_3[\mathrm{cos}(k_x a_L)+\mathrm{cos}(k_y a_L)]\bm{\sigma}_y.
\end{align}
When $\Delta$ and $\theta_3$ are small, we get (see Appendix \ref{AppendixC2})
\begin{align}
\label{eqn63}
m=\frac{1}{2(t_1+t_3)a_L^2},\quad &M_d=\frac{\Delta^2}{(t_2+4t_3)^2},\nonumber\\
M''=\frac{8}{3}W=\frac{4a^2_Lt_3\Delta^2}{(t_2+4t_3)^3},\quad &S=-\frac{a^2_Lt_2t_3\theta_3\Delta}{(t_2+4t_3)^3}.
\end{align}

The Bloch wavefunction quantities ($M_d$, $M''$, $W$, $S$) are proportional to $\Delta^2$ or $\theta_3\Delta$, so they vanish when the sublattice symmetry is restored. This is reasonable because they characterize sublattice structures of Bloch wavefunctions. $S$ is proportional to $\theta_3$ as it is time-reversal odd. Moreover, the quantum geometric tensor $\mathcal{B}=\mathsf{g}+i\Omega=0$ at $\bm{k}=\bm{0}$, because it only contains first-order derivatives of wavefunctions. Although $\mathcal{B}=0$, there are still lattice-induced wavefunction effects, which are clearly beyond the traditional description.

We can estimate orders of magnitude of the Bloch wavefunction quantities and resulting observables in typical experimental setups of cold atoms (e.g.~$\mathrm{Rb}^{87}$). Typical wavelength and strength of lasers, chemical potential of atoms, and harmonic confinement~\cite{Bloch2005, Jo2012, Tarruell2012, Jotzu2014, Taie2015, Aycock2017} give $a_L\sim 1\mathrm{\mu m}$, $t_{1,2,3}\sim\mu\sim 1\mathrm{kHz}$, $\omega_0\sim 10\mathrm{Hz}$. The condition $\omega_0\ll\mu\sim t_{1,2,3}$ is satisfied, so the gradient expansion applies to the hydrodynamic theory. To evaluate $\Delta$ and $t_3\theta_3$, we take them at a typical order of a mass breaking sublattice symmetry or time-reversal symmetry~\cite{Jotzu2014}, $\Delta\sim t_3\theta_3\sim 100\mathrm{Hz}$. Then Eq.~(\ref{eqn63}) leads to $M_d\sim m\mu M''\sim m\mu S\sim 0.01$. According to Eqs.~(\ref{eqn48}, \ref{eqn59}, \ref{eqn60}), experimental obervables are obtained, $\delta\omega_{1,0}\sim\mathrm{0.1Hz}$, $\mathsf{b}_{1,0}/\bar{n}_0\sim 10^{-2}$, $\tau_{1,0}\sim 1\mathrm{ms}$. Here $\delta\omega_{1,0}$ is the lattice correction to $\omega_{1,0}$, $\bar{n}_0\sim\mu/g$ is the average density of the atoms, and the relative amplitude of density $\mathsf{b}_{1,0}/\bar{n}_0$ can be detected by absorption strength of light~\cite{Anderson1995, Davis1995}. In this example, the lattice corrections are about \%1, which are small but may be in the same order as the quantum-fluctuation corrections discussed in Sec.~\ref{Sec34}. In this case, the two kinds of corrections should be added together.

\section{\label{Sec6}Summary and outlook}

In this paper, we study lattice-induced wavefunction effects on correlations between bosonic particles in lattices. The hydrodynamic properties of superfluids are influenced by a richer structure of Bloch wavefunctions beyond the single-body quantum geometric tensor. The lattice-induced wavefunction effects can be observed by the dynamic distribution of the particle density.

Our discussions are restricted to the simplest quantum many-body system where the interaction is local and the resulting effective theory is isotropic. The author wishes this paper to promote further exploration and application of wavefunction effects in wider many-body systems. 

Besides, the discussions are still based on single-body Bloch wavefunctions. We derive the results at a practical level. Namely, this paper aims to propose specific many-body physical effects of Bloch wavefunctions, instead of to clarify more systematic or mathematical structures of two-body or many-body quantum states. Thus, a better understanding of quantum geometry in many-body systems and corresponding physical consequences remains intended~\cite{Hetenyi2023, Salerno2023, Torma2023}.

\begin{acknowledgments}

The author is grateful to Ryuichi Shindou for his fruitful suggestions and support. The author also thanks Lingxian Kong, Zhenyu Xiao, Xuesong Hu, Pengwei Zhao, and Siyu Pan for their helpful discussions. The work was supported by the National Basic Research Programs of China (No.~2019YFA0308401) and the National Natural Science Foundation of China (No.~11674011 and No.~12074008).\\

\end{acknowledgments}

\appendix

\section{\label{AppendixA}Derivation of the second-order effective interaction}

In this Appendix, we derive the second-order effective interaction (Eqs.~(\ref{eqn15}--\ref{eqn17})) in Subsection \ref{AppendixA1} and the form after the gauge transformation (Eq.~(\ref{eqn18})) in Subsection \ref{AppendixA2}. We denote\begin{align}
\label{eqnA1}
\frac{2}{g}H_{I2}\equiv&\frac{1}{(2\pi)^d}\int_\infty(\prod_{i=1}^3 d^d\bm{k}_i)b^\dagger_{\bm{k}_1}b^\dagger_{\bm{k}_2}b_{\bm{k}_3}b_{\bm{k}_4}F_2(\bm{k}_1,\bm{k}_2,\bm{k}_3,\bm{k}_4).
\end{align}
We define three independent momenta respectively representing the Cooper, direct, and exchange channels,
\begin{align}
\label{eqnA2}
\bm{s}\equiv\bm{k}_1+\bm{k}_2,\quad\bm{t}\equiv\bm{k}_3-\bm{k}_1,\quad\bm{u}\equiv\bm{k}_3-\bm{k}_2.
\end{align}

\subsection{\label{AppendixA1}Momentum expansion and Fourier transorm}

Second-order terms in $\langle\langle u_{\bm{k}_1}u_{\bm{k}_2}|u_{\bm{k}_3}u_{\bm{k}_4}\rangle\rangle$ read
\begin{align}
\label{eqnA3}
&F_2(\bm{k}_1,\bm{k}_2,\bm{k}_3,\bm{k}_4)=\frac{1}{2}\langle\langle u\partial_{k_\alpha} \partial_{k_\beta} u|uu\rangle\rangle_{\bm{0}}(k_{1\alpha}k_{1\beta}+k_{2\alpha}k_{2\beta})\nonumber\\
&\!\ \!\ \!\ \!\ +\frac{1}{2}\langle\langle uu|u\partial_{k_\alpha}\partial_{k_\beta} u\rangle\rangle_{\bm{0}}(k_{3\alpha}k_{3\beta}+k_{4\alpha}k_{4\beta})\nonumber\\
&\!\ \!\ \!\ \!\ +\langle\langle(\partial_{k_\alpha} u)(\partial_{k_\beta} u)|uu\rangle\rangle_{\bm{0}}k_{1\alpha}k_{2\beta}\nonumber\\
&\!\ \!\ \!\ \!\ +\langle\langle uu|(\partial_{k_\alpha} u)(\partial_{k_\beta} u)\rangle\rangle_{\bm{0}}k_{3\alpha}k_{4\beta}\nonumber\\
&\!\ \!\ \!\ \!\ +\langle\langle u\partial_{k_\alpha} u|u\partial_{k_\beta} u\rangle\rangle_{\bm{0}}(k_{1\alpha}+k_{2\alpha})(k_{3\beta}+k_{4\beta}).
\end{align}
With momentum conservation, we get
\begin{align}
\label{eqnA4}
&F_2(\bm{k}_1,\bm{k}_2,\bm{k}_3,\bm{k}_4)|_{\bm{k}_4=\bm{k}_1+\bm{k}_2-\bm{k}_3}\nonumber\\
=&\frac{1}{4}(\langle\langle uu|(\partial_{k_\alpha} u)(\partial_{k_\beta} u)\rangle\rangle_{\bm{0}}+\langle\langle uu|u\partial_{k_\alpha}\partial_{k_\beta} u\rangle\rangle_{\bm{0}}\nonumber\\
&+\langle\langle u\partial_{k_\alpha} u|u\partial_{k_\beta} u\rangle\rangle_{\bm{0}}+\langle\langle u\partial_{k_\beta} u|u\partial_{k_\alpha} u\rangle\rangle_{\bm{0}}+\mathrm{c.c.})s_\alpha s_\beta\nonumber\\
&+\frac{1}{4}(\langle\langle uu|u\partial_{k_\alpha} \partial_{k_\beta} u\rangle\rangle_{\bm{0}}-\langle\langle uu|(\partial_{k_\alpha} u)(\partial_{k_\beta} u)\rangle\rangle_{\bm{0}}\nonumber\\
&\!\ \!\ \!\ \!\ +\mathrm{c.c.})(t_\alpha t_\beta+u_\alpha u_\beta)\nonumber\\
&+\frac{1}{4}(\langle\langle uu|u\partial_{k_\alpha} \partial_{k_\beta} u\rangle\rangle_{\bm{0}}-\langle\langle uu|(\partial_{k_\alpha} u)(\partial_{k_\beta} u)\rangle\rangle_{\bm{0}}\nonumber\\
&\!\ \!\ \!\ \!\ -\mathrm{c.c.})(t_\alpha u_\beta+u_\alpha t_\beta),
\end{align}
where we have used
\begin{align}
\label{eqnA5}
\bm{k}_1-\bm{k}_2=(\bm{k}_3-\bm{k}_2)-(\bm{k}_3-\bm{k}_1),
\end{align}
\begin{align}
\label{eqnA6}
\bm{k}_3-\bm{k}_4=&(\bm{k}_3-\bm{k}_1)+(\bm{k}_1-\bm{k}_4)\nonumber\\
=&(\bm{k}_3-\bm{k}_1)+(\bm{k}_3-\bm{k}_2).
\end{align}
Note that the direct channel and the exchange channel are symmetric.

Because $\langle\langle uu|uu\rangle\rangle_{\bm{k}}\equiv\langle\langle u_{\bm{k}}u_{\bm{k}}|u_{\bm{k}}u_{\bm{k}}\rangle\rangle$ is not a constant, we have two gauge-independent quantities characterizing momentum dependence,
\begin{align}
\label{eqnA7}
\partial_{k_\alpha}\langle\langle uu|uu\rangle\rangle_{\bm{k}}=2\langle\langle u\partial_{k_\alpha} u|uu\rangle\rangle_{\bm{k}}+2\langle\langle uu|u\partial_{k_\alpha} u\rangle\rangle_{\bm{k}},
\end{align}
\begin{align}
\label{eqnA8}
&\partial_{k_\alpha}\partial_{k_\beta}\langle\langle uu|uu\rangle\rangle_{\bm{k}}\nonumber\\
=&4\langle\langle u\partial_{k_\alpha} u|u\partial_{k_\beta} u\rangle\rangle_{\bm{k}}+2\langle\langle (\partial_{k_\alpha} u)(\partial_{k_\beta} u)|u u\rangle\rangle_{\bm{k}}\nonumber\\
&+2\langle\langle u\partial_{k_\alpha}\partial_{k_\beta} u|uu\rangle\rangle_{\bm{k}}+4\langle\langle u\partial_{k_\beta} u|u\partial_{k_\alpha} u\rangle\rangle_{\bm{k}}\nonumber\\
&+2\langle\langle uu|(\partial_{k_\alpha} u)(\partial_{k_\beta} u)\rangle\rangle_{\bm{k}}+2\langle\langle uu|u\partial_{k_\alpha}\partial_{k_\beta} u\rangle\rangle_{\bm{k}}.
\end{align}
Besides, we can apply derivatives on $|u_{\bm{k}}|^2$ to construct a gauge-independent quantity,
\begin{align}
\label{eqnA9}
&\langle\langle \partial_{k_\alpha}(u^*u)|\partial_{k_\beta}(u^*u)\rangle\rangle_{\bm{k}}\nonumber\\
\equiv&\frac{(2\pi)^{2d}}{\Omega_{\mathrm{cell}}}\int d^d\bm{r}\partial_{k_\alpha} [u_{\bm{k}}^*(\bm{r})u_{\bm{k}}(\bm{r})]\partial_{k_\beta} [u_{\bm{k}}^*(\bm{r})u_{\bm{k}}(\bm{r})]\nonumber\\
=&\langle\langle u\partial_{k_\alpha} u|u \partial_{k_\beta} u\rangle\rangle_{\bm{k}}+\langle\langle u\partial_{k_\beta} u|u \partial_{k_\alpha} u\rangle\rangle_{\bm{k}}\nonumber\\
&+\langle\langle (\partial_{k_\alpha} u)(\partial_{k_\beta} u)|uu\rangle\rangle_{\bm{k}}+\langle\langle uu|(\partial_{k_\alpha} u)(\partial_{k_\beta} u)\rangle\rangle_{\bm{k}}.
\end{align}
Then two of the terms in Eq.~(\ref{eqnA4}) are gauge-independent,
\begin{align}
\label{eqnA10}
&F_2(\bm{k}_1,\bm{k}_2,\bm{k}_3,\bm{k}_4)|_{\bm{k}_4=\bm{k}_1+\bm{k}_2-\bm{k}_3}\nonumber\\
=&\frac{1}{8}M''_{\alpha\beta}(\bm{0})s_\alpha s_\beta+\frac{1}{8}[M''_{\alpha\beta}(\bm{0})-4P_{\alpha\beta}(\bm{0})](t_\alpha t_\beta+u_\alpha u_\beta)\nonumber\\
&+\frac{i}{2}\widetilde{S}_{\alpha\beta}(\bm{0})(t_\alpha u_\beta+u_\alpha t_\beta).
\end{align}

We can further express $H_{I2}$ in the real space with the hydrodynamic variables. With
\begin{align}
\label{eqnA11}
b(\bm{r})=\int_{\infty}\frac{d^d\bm{k}}{(2\pi)^{\frac{d}{2}}}b_{\bm{k}}e^{i\bm{k}\cdot\bm{r}},
\end{align}
we get
\begin{align}
\label{eqnA12}
&\frac{1}{(2\pi)^d}\int_\infty(\prod_{i=1}^3 d^d\bm{k}_i)b^\dagger_{\bm{k}_1}b^\dagger_{\bm{k}_2}b_{\bm{k}_3}b_{\bm{k}_4}\frac{1}{8}M''_{\alpha\beta}(\bm{0})s_\alpha s_\beta\nonumber\\
=&\int_\infty d^d\bm{r}\frac{1}{8}M''_{\alpha\beta}(\bm{0})[\partial_\alpha (b^\dagger b^\dagger)][\partial_\beta (b b)]\nonumber\\
=&\int_\infty d^d\bm{r}\frac{1}{8}M''_{\alpha\beta}(\bm{0})[(\partial_\alpha n)(\partial_\beta n)+4n^2(\partial_\alpha \theta)(\partial_\beta \theta)],
\end{align}
\begin{align}
\label{eqnA13}
&\frac{1}{(2\pi)^d}\int_\infty(\prod_{i=1}^3 d^d\bm{k}_i)b^\dagger_{\bm{k}_1}b^\dagger_{\bm{k}_2}b_{\bm{k}_3}b_{\bm{k}_4}\frac{1}{8}[M''_{\alpha\beta}(\bm{0})\nonumber\\
&\!\ \!\ \!\ \!\ -4P_{\alpha\beta}(\bm{0})](t_\alpha t_\beta+u_\alpha u_\beta)\nonumber\\
=&\int_\infty d^d\bm{r}\frac{1}{4}[M''_{\alpha\beta}(\bm{0})-4P_{\alpha\beta}(\bm{0})][\partial_\alpha (b^\dagger b)][\partial_\beta (b^\dagger b)]\nonumber\\
=&\int_\infty d^d\bm{r}[\frac{1}{4}M''_{\alpha\beta}(\bm{0})-P_{\alpha\beta}(\bm{0})](\partial_\alpha n)(\partial_\beta n),
\end{align}
\begin{align}
\label{eqnA14}
&\frac{1}{(2\pi)^d}\int_\infty(\prod_{i=1}^3 d^d\bm{k}_i)b^\dagger_{\bm{k}_1}b^\dagger_{\bm{k}_2}b_{\bm{k}_3}b_{\bm{k}_4}[\frac{i}{2}\widetilde{S}_{\alpha\beta}(\bm{0})](t_\alpha u_\beta+u_\alpha t_\beta)\nonumber\\
=&\int_\infty d^d\bm{r}\widetilde{S}_{\alpha\beta}(\bm{0})\frac{i}{2}[b^\dagger(\partial_\alpha\partial_\beta b^\dagger)b b-(\partial_\alpha b^\dagger)(\partial_\beta b^\dagger)b b\nonumber\\
&\!\ \!\ \!\ \!\ -b^\dagger b^\dagger b(\partial_\alpha\partial_\beta b)+b^\dagger b^\dagger(\partial_\alpha b)(\partial_\beta b)]\nonumber\\
=&-2\int_\infty d^d\bm{r}\widetilde{S}_{\alpha\beta}(\bm{0})n(\partial_\alpha n)(\partial_\beta\theta).
\end{align}
When deriving Eq.~(\ref{eqnA14}), we have used
\begin{align}
\label{eqnA15}
&\frac{1}{2}(t_\alpha u_\beta+u_\alpha t_\beta)\nonumber\\
=&-\frac{1}{4}(k_{1\alpha}-k_{2\alpha})(k_{1\beta}-k_{2\beta})-(k_{3\alpha}-k_{4\alpha})(k_{3\beta}-k_{4\beta})]\nonumber\\
=&-\frac{1}{2}(k_{1\alpha}k_{1\beta}-k_{1\alpha}k_{2\beta}-k_{3\alpha}k_{3\beta}+k_{3\alpha}k_{4\beta}),
\end{align}
\begin{align}
\label{eqnA16}
&b^\dagger(\partial_\alpha\partial_\beta b^\dagger)b b-(\partial_\alpha b^\dagger)(\partial_\beta b^\dagger)b b\nonumber\\
=&n^{\frac{3}{2}}\partial_\alpha \partial_\beta n-n(\partial_\alpha n^{\frac{1}{2}})(\partial_\beta n^{\frac{1}{2}})-in^2\partial_\alpha\partial_\beta\theta.
\end{align}
Eqs.~(\ref{eqnA1}, \ref{eqnA10}, \ref{eqnA12}--\ref{eqnA14}) are consistent with Eqs.~(\ref{eqn15}--\ref{eqn17}).

\subsection{\label{AppendixA2}Gauge transformation}

Under the gauge transformation,
\begin{align}
\label{eqnA17}
u_{\bm{k}}\rightarrow\tilde{u}_{\bm{k}}\equiv u_{\bm{k}}e^{i\vartheta_{\bm{k}}},
\end{align}
$\widetilde{S}(\bm{k})$ is changed,
\begin{align}
\label{eqnA18}
&\langle\langle\tilde{u}\tilde{u}|\tilde{u}\partial_{k_\alpha}\partial_{k_\beta} \tilde{u}\rangle\rangle_{\bm{k}}-\langle\langle\tilde{u}\tilde{u}|(\partial_{k_\alpha}\tilde{u})(\partial_{k_\beta}\tilde{u})\rangle\rangle_{\bm{k}}-\mathrm{c.c.}\nonumber\\
=&\langle\langle uu|ue^{-i\vartheta}\partial_{k_\alpha}[(\partial_{k_\beta} u)e^{i\vartheta}+i(\partial_{k_\beta}\vartheta)ue^{i\vartheta}]\rangle\rangle_{\bm{k}}\nonumber\\
&-\langle\langle uu|[\partial_{k_\alpha} u+i(\partial_{k_\alpha}\vartheta)u][\partial_{k_\beta} u+i(\partial_{k_\beta}\vartheta)u]\rangle\rangle_{\bm{k}}-\mathrm{c.c.}\nonumber\\
=&\langle\langle uu|u\partial_{k_\alpha}\partial_{k_\beta} u\rangle\rangle_{\bm{k}}-\langle\langle uu|(\partial_{k_\alpha} u)(\partial_{k_\beta} u)\rangle\rangle_{\bm{k}}\nonumber\\
&+i(\partial_{k_\alpha}\partial_{k_\beta}\vartheta)\langle \langle uu|uu\rangle\rangle_{\bm{k}}-\mathrm{c.c.}.
\end{align}
With the gauge Eq.~(\ref{eqn8}), we get
\begin{align}
\label{eqnA19}
&\langle\langle\tilde{u}\tilde{u}|\tilde{u}\partial_{k_\alpha}\partial_{k_\beta} \tilde{u}\rangle\rangle_{\bm{0}}-\langle\langle\tilde{u}\tilde{u}|(\partial_{k_\alpha}\tilde{u})(\partial_{k_\beta}\tilde{u})\rangle\rangle_{\bm{0}}-\mathrm{c.c.}\nonumber\\
=&\langle\langle uu|u\partial_{k_\alpha}\partial_{k_\beta} u\rangle_{\bm{0}}-\langle\langle uu|(\partial_{k_\alpha} u)(\partial_{k_\beta} u)\rangle\rangle_{\bm{0}}\nonumber\\
&+\frac{i}{2}\langle\langle uu|uu\rangle\rangle_{\bm{0}}[\partial_{k_\alpha} A_\beta(\bm{0})+\partial_{k_\beta} A_\alpha(\bm{0})]-\mathrm{c.c.}\nonumber\\
=&\langle\langle uu|u\partial_{k_\alpha}\partial_{k_\beta} u\rangle\rangle_{\bm{0}}-\langle\langle uu|(\partial_{k_\alpha} u)(\partial_{k_\beta} u)\rangle\rangle_{\bm{0}}\nonumber\\
&-\frac{1}{2}\langle\langle uu|uu\rangle\rangle_{\bm{0}}(\partial_{k_\alpha}\langle u|\partial_{k_\beta} u\rangle_{\bm{0}}+\partial_{k_\beta}\langle u|\partial_{k_\alpha} u\rangle_{\bm{0}})-\mathrm{c.c.}\nonumber\\
=&\langle\langle uu|u\partial_{k_\alpha}\partial_{k_\beta} u\rangle\rangle_{\bm{0}}-\langle\langle uu|(\partial_{k_\alpha} u)(\partial_{k_\beta} u)\rangle\rangle_{\bm{0}}\nonumber\\
&-\langle\langle uu|uu\rangle\rangle_{\bm{0}}\langle u|\partial_{k_\alpha}\partial_{k_\beta} u\rangle_{\bm{0}}-\mathrm{c.c.}.
\end{align}
So $\widetilde{S}_{\alpha\beta}(\bm{0})$ in Eq.~(\ref{eqn15}) should be substituted by $S_{\alpha\beta}(\bm{0})$ defined in Eq.~(\ref{eqn18}). We can also verify that $S_{\alpha\beta}(\bm{k})$ is gauge-independent,
\begin{align}
\label{eqnA20}
&\langle\langle \tilde{u}\tilde{u}|\tilde{u}\partial_{k_\alpha}\partial_{k_\beta} \tilde{u}\rangle\rangle_{\bm{k}}-\langle\langle \tilde{u}\tilde{u}|(\partial_{k_\alpha}\tilde{u})(\partial_{k_\beta} \tilde{u})\rangle\rangle_{\bm{k}}\nonumber\\
&+\frac{i}{2}\langle\langle \tilde{u}\tilde{u}|\tilde{u}\tilde{u}\rangle\rangle_{\bm{k}} [\partial_\alpha \tilde{A}_\beta(\bm{k})+\partial_\beta \tilde{A}_\alpha(\bm{k})]\nonumber\\
=&\langle\langle uu|u\partial_{k_\alpha}\partial_{k_\beta} u\rangle\rangle_{\bm{k}}-\langle\langle uu|\partial_{k_\alpha} u\partial_{k_\beta} u\rangle\rangle_{\bm{k}}\nonumber\\
&+i\partial_{k_\alpha}\partial_{k_\beta}\vartheta(\bm{k})\langle \langle uu|uu\rangle\rangle_{\bm{k}}+\frac{i}{2}\langle\langle uu|uu\rangle\rangle_{\bm{k}}[\partial_{k_\alpha} A_\beta(\bm{k})\nonumber\\
&\!\ \!\ \!\ \!\ -\partial_{k_\alpha}\partial_{k_\beta}\vartheta(\bm{k})+\partial_{k_\beta} A_\alpha(\bm{k})-\partial_{k_\alpha}\partial_{k_\beta}\vartheta(\bm{k})]\nonumber\\
=&\langle\langle uu|u\partial_{k_\alpha}\partial_{k_\beta} u\rangle\rangle_{\bm{k}}-\langle\langle uu|\partial_{k_\alpha} u\partial_{k_\beta} u\rangle\rangle_{\bm{k}}\nonumber\\
&+\frac{i}{2}\langle\langle uu|uu\rangle\rangle_{\bm{k}}[\partial_{k_\alpha} A_\beta(\bm{k})+\partial_{k_\beta} A_\alpha(\bm{k})],
\end{align}
where $\tilde{A}_\alpha(\bm{k})\equiv i\langle\tilde{u}|\partial_{k_\alpha}\tilde{u}\rangle_{\bm{k}}$.

\section{\label{AppendixB}Solutions of collective modes in a harmonic trap}

In this Appendix, we solve the linear homogeneous ordinary differential equations Eqs.~(\ref{eqn37}, \ref{eqn39}) with proper boundary conditions to get the low-energy collective modes in the harmonic trap. We first take $\eta=0$ to review the results without the optical lattice in Subsection \ref{AppendixB1}. The general case of $\eta\neq 0$ is solved in Subsection \ref{AppendixB2}. Obvious differences can be seen by a comparison of the two cases. Besides, we prove the orthonormal relation Eq.~(\ref{eqn42}) in Subsection \ref{AppendixB3}. We omit the subscripts of $\rho_{\mathsf{n},\mathsf{j}}$ and $\omega_{\mathsf{n},\mathsf{j}}$ when there is no ambiguity and denote $\kappa=\frac{\omega^2}{\omega_0^2}$.

\subsection{\label{AppendixB1}Without lattice-induced wavefunction effects}

\subsubsection{\label{AppendixB11}3D case}

When $\eta=0$, Eq.~(\ref{eqn37}) becomes
\begin{align}
\label{eqnB1}
&\frac{1}{2}(1-r^2)\frac{d^2\rho}{d r^2}+(1-2r^2)\frac{1}{r}\frac{d\rho}{d r}\nonumber\\
&+[\kappa-(1-r^2)\frac{\ell(\ell+1)}{2r^2}]\rho=0.
\end{align}
When $r\rightarrow 0$, we get
\begin{align}
\label{eqnB2}
\frac{1}{2}\frac{d^2\rho}{d r^2}+\frac{1}{r}\frac{d\rho}{d r}-\frac{\ell(\ell+1)}{2r^2}\rho=0.
\end{align}
Taking $\rho=r^l$ into Eq.~(\ref{eqnB2}), we get $l(l+1)=\ell(\ell+1)$, so $l=\ell$ or $l=-\ell-1$. Because $\ell\in\mathbb{N}$, to make $\rho(r)$ finite when $r\rightarrow 0$, we take $l=\ell$. Suppose an eigenfunction is a polynomial,
\begin{align}
\label{eqnB3}
\rho(r)=\sum_{\mathsf{k}'=0}^{\mathsf{N}'} \alpha_{\mathsf{k}'}r^{\mathsf{k}'+\ell},
\end{align}
and get
\begin{align}
\label{eqnB4}
&\sum_{\mathsf{k}'=-2}^{\mathsf{N}'-2}\alpha_{\mathsf{k}'+2}[\frac{1}{2}(\ell+\mathsf{k}'+2)(\ell+\mathsf{k}'+1)\nonumber\\
&\!\ \!\ \!\ \!\ +(\ell+\mathsf{k}'+2)-\frac{1}{2}\ell(\ell+1)]r^{\ell+\mathsf{k}'}\nonumber\\
=&\sum_{\mathsf{k}=0}^{\mathsf{N}'}\alpha_{\mathsf{k}'}[\frac{1}{2}(\ell+\mathsf{k}')(\ell+\mathsf{k}'-1)\nonumber\\
&\!\ \!\ \!\ \!\ +2(\ell+\mathsf{k}')-\kappa-\frac{1}{2}\ell(\ell+1)]r^{\ell+\mathsf{k}'}.
\end{align}
An equation for $r^{\ell-2}$ is automatically satisfied because it is equivalent to the asymptotic equation Eq.~(\ref{eqnB2}). Equations for $r^{\ell+2\mathsf{k}'-1}$ ($\mathsf{k}'\in\mathbb{N}$) requires $\mathsf{k}'$ and $\mathsf{N}'$ to be even numbers, so we can denote $\mathsf{k}'=2\mathsf{k}$, $\mathsf{N}'=2\mathsf{N}$. Equations for $r^{\ell+2\mathsf{k}}$ ($0<\mathsf{k}<\mathsf{N}$, $\mathsf{k}\in\mathbb{N}$) give a recursion condition,
\begin{align}
\label{eqnB5}
&\alpha_{2\mathsf{k}+2}[\frac{1}{2}(\ell+2\mathsf{k}+2)(\ell+2\mathsf{k}+1)+(\ell+2\mathsf{k}+2)\nonumber\\
&\!\ \!\ -\frac{1}{2}\ell(\ell+1)]=\alpha_{2\mathsf{k}}[\frac{1}{2}(\ell+2\mathsf{k})(\ell+2\mathsf{k}-1)\nonumber\\
&\!\ \!\ \!\ \!\ +2(\ell+2\mathsf{k})-\kappa-\frac{1}{2}\ell(\ell+1)].
\end{align}
When $\mathsf{N}$ is finite, $\mathsf{N}=\mathsf{n}$ and an equation for $r^{\ell+2\mathsf{n}}$ gives
\begin{align}
\label{eqnB6}
\kappa=&\frac{1}{2}(\ell+2\mathsf{n})(\ell+2\mathsf{n}-1)+2(\ell+2\mathsf{n})-\frac{1}{2}\ell(\ell+1)\nonumber\\
=&2\mathsf{n}^2+2\mathsf{n}\ell+3\mathsf{n}+\ell.
\end{align}
Taking Eq.~(\ref{eqnB6}) into Eq.~(\ref{eqnB5}), we have
\begin{align}
\label{eqnB7}
\alpha_{2\mathsf{k}+2}=-\frac{(\mathsf{n}-\mathsf{k})(2\mathsf{k}+2\mathsf{n}+2\ell+3)}{(\mathsf{k}+1)(2\mathsf{k}+2\ell+3)}\alpha_{2\mathsf{k}}.
\end{align}

Below we explain why we cannot take $\mathsf{N}=\infty$. In this case, when $\mathsf{k}\rightarrow\infty$, Eq.~(\ref{eqnB5}) gives
\begin{align}
\label{eqnB8}
\frac{\alpha_{2\mathsf{k}+2}}{\alpha_{2\mathsf{k}}}=\frac{2+\frac{2\ell+3}{\mathsf{k}}}{2+\frac{2\ell+5}{\mathsf{k}}}+\mathcal{O}(\frac{1}{\mathsf{k}^2})=1-\frac{1}{\mathsf{k}}+\mathcal{O}(\frac{1}{\mathsf{k}^2}).
\end{align}
We can take a test form of $\alpha_{2\mathsf{k}}$ for $\mathsf{k}\rightarrow\infty$,
\begin{align}
\label{eqnB9}
\alpha_{2\mathsf{k}}=\frac{C}{\mathsf{k}^\zeta}(1+\frac{C'}{\mathsf{k}^{\zeta'}}),
\end{align}
where $\zeta>0$, $\zeta'>0$, $C$ and $C'$ are constant, then
\begin{align}
\label{eqnB10}
&\frac{\alpha_{2\mathsf{k}+2}}{\alpha_{2\mathsf{k}}}=\frac{\mathsf{k}^\zeta}{(\mathsf{k}+1)^\zeta}\frac{1+\frac{C'}{(\mathsf{k}+1)^{\zeta'}}}{1+\frac{C'}{\mathsf{k}^{\zeta'}}}\nonumber\\
=&(1-\frac{\zeta}{\mathsf{k}})[1+\frac{C'}{(\mathsf{k}+1)^{\zeta'}}](1-\frac{C'}{\mathsf{k}^{\zeta'}})+\mathcal{O}(\frac{1}{\mathsf{k}^{\zeta'+1}})+\mathcal{O}(\frac{1}{\mathsf{k}^2})\nonumber\\
=&1-\frac{\zeta}{\mathsf{k}}+\mathcal{O}(\frac{1}{\mathsf{k}^{\zeta'+1}})+\mathcal{O}(\frac{1}{\mathsf{k}^2}).
\end{align}
Comparing Eq.~(\ref{eqnB10}) with Eq.~(\ref{eqnB8}), we get $\zeta=1$. Eq.~(\ref{eqnB9}) can be generalized to show that $\alpha_{2\mathsf{k}}=\frac{C}{\mathsf{k}}+\mathcal{O}(\frac{1}{\mathsf{k}^2})$. Then we get
\begin{align}
\label{eqnB11}
\rho(r\rightarrow 1)=\sum_{\mathsf{k}=0}^{\mathsf{k}_0}\alpha_{2\mathsf{k}}+\sum_{\mathsf{k}=\mathsf{k}_0}^\infty[\frac{C}{\mathsf{k}}+\mathcal{O}(\frac{1}{\mathsf{k}^2})]\rightarrow\infty,
\end{align}
where $\mathsf{k}_0$ is a constant large enough. It is unphysical to have a divergent $\rho_{\mathsf{n},\ell}$ when $r\rightarrow 1$. Note that at least one of $\mathsf{n}$ and $\ell$ is nonzero in the divergent case, so $n(\bm{r},t)$ becomes negative for some spacetime coordinates. To avoid the divergence, we need Eq.~(\ref{eqnB6}) to truncate the sequence $\alpha_{2k}$ and get a finite $\mathsf{N}$. 

In summary, eigenfunctions take a form,
\begin{align}
\label{eqnB12}
\rho_{\mathsf{n},\ell}=\sum_{\mathsf{k}=0}^{\mathsf{N}} \alpha_{\mathsf{n},\ell,2\mathsf{k}}r^{2\mathsf{k}+\ell},
\end{align}
where $\alpha_{\mathsf{n},\ell,2\mathsf{k}}$ satisfies Eq.~(\ref{eqnB7}). When $\eta=0$, we get $\mathsf{N}=\mathsf{n}$ in Eq.~(\ref{eqnB12}), but we will show that this is not true for $\eta\neq 0$.

\subsubsection{\label{AppendixB11}2D case}

When $\eta=0$, Eq.~(\ref{eqn39}) becomes
\begin{align}
\label{eqnB13}
&\frac{1}{2}(1-r^2)\frac{d^2\rho}{d r^2}+\frac{1}{2}(1-3r^2)\frac{1}{r}\frac{d\rho}{d r}\nonumber\\
&+[\kappa r^2-\frac{1}{2}m^2(1-r^2)]\frac{\rho}{r^2}=0.
\end{align}
An analysis of $r\rightarrow 0$ gives $\rho(r\rightarrow 0)\rightarrow r^{|\mathsf{m}|}$, then we get
\begin{align}
\label{eqnB14}
\rho_{\mathsf{n},|\mathsf{m}|}=\sum_{\mathsf{k}=0}^{\mathsf{N}} \alpha_{\mathsf{n},|\mathsf{m}|,2\mathsf{k}}r^{2\mathsf{k}+|\mathsf{m}|}.
\end{align}
Let us take $\mathsf{m}\geq 0$ for simplicity. There is a recursion relation,
\begin{align}
\label{eqnB15}
&\alpha_{2\mathsf{k}+2}[\frac{1}{2}(\mathsf{m}+2\mathsf{k}+2)(\mathsf{m}+2\mathsf{k}+1)+\frac{1}{2}(\mathsf{m}+2\mathsf{k}+2)\nonumber\\
&\!\ \!\ -\frac{1}{2}\mathsf{m}^2]=\alpha_{2\mathsf{k}}[\frac{1}{2}(\mathsf{m}+2\mathsf{k})(\mathsf{m}+2\mathsf{k}-1)\nonumber\\
&\!\ \!\ \!\ \!\ +\frac{3}{2}(\mathsf{m}+2\mathsf{k})-\kappa-\frac{1}{2}\mathsf{m}^2].
\end{align}
If $\mathsf{N}$ is finite, we get $\mathsf{N}=\mathsf{n}$ and
\begin{align}
\label{eqnB16}
\kappa=&\frac{1}{2}(\mathsf{m}+2\mathsf{n})(\mathsf{m}+2\mathsf{n}-1)+\frac{3}{2}(\mathsf{m}+2\mathsf{n})-\frac{1}{2}\mathsf{m}^2\nonumber\\
=&2\mathsf{n}^2+2\mathsf{n}\mathsf{m}+2\mathsf{n}+\mathsf{m},
\end{align}
\begin{align}
\label{eqnB17}
\alpha_{2\mathsf{k}+2}=&\alpha_{2\mathsf{k}}\frac{2(\mathsf{k}-\mathsf{n})(\mathsf{k}+\mathsf{n})+2(\mathsf{k}-\mathsf{n})(\mathsf{m}+1)}{\frac{1}{2}(\mathsf{m}+2\mathsf{k}+2)^2-\frac{1}{2}\mathsf{m}^2}\nonumber\\
=&\alpha_{2\mathsf{k}}\frac{(\mathsf{k}-\mathsf{n})(\mathsf{k}+\mathsf{n}+\mathsf{m}+1)}{(\mathsf{k}+1)(\mathsf{k}+\mathsf{m}+1)}.
\end{align}
If $\mathsf{N}=\infty$, then when $\mathsf{k}\rightarrow\infty$, we get
\begin{align}
\label{eqnB18}
\frac{\alpha_{2\mathsf{k}+2}}{\alpha_{2\mathsf{k}}}=\frac{2\mathsf{k}^2+(2\mathsf{m}+2)\mathsf{k}}{2\mathsf{k}^2+(2\mathsf{m}+4)\mathsf{k}}+\mathcal{O}(\frac{1}{\mathsf{k}^2})=1-\frac{1}{\mathsf{k}}+\mathcal{O}(\frac{1}{\mathsf{k}^2}),
\end{align}
and by the analysis below Eq.~(\ref{eqnB8}) we know it is impossible.

\subsection{\label{AppendixB2}With lattice-induced wavefunction effects}

\subsubsection{\label{AppendixB21}3D case}

Consider Eq.~(\ref{eqn37}) with $\eta\neq 0$. Note that $\eta$ can be an $\mathcal{O}(1)$ quantity. We have an equation,
\begin{align}
\label{eqnB19}
&\frac{1}{2}[(1+\frac{\eta}{2})-(1+\eta)r^2+\frac{\eta}{2}r^4]\frac{d^2\rho}{d r^2}\nonumber\\
&+[(1+\frac{\eta}{2})-2(1+\eta)r^2+\frac{3\eta}{2}r^4]\frac{1}{r}\frac{d\rho}{d r}\nonumber\\
&+\{\kappa r^2-\frac{1}{2}\ell(\ell+1)[(1+\frac{\eta}{2})-(1+\eta)r^2+\frac{\eta}{2}r^4]\}\frac{\rho}{r^2}=0.
\end{align}
When $r\rightarrow 0$, we get
\begin{align}
\label{eqnB20}
\frac{1}{2}(1+\frac{\eta}{2})\frac{d^2\rho}{d r^2}+(1+\frac{\eta}{2})\frac{1}{r}\frac{d\rho}{d r}-\frac{1}{2}\ell(\ell+1)(1+\frac{\eta}{2})\frac{\rho}{r^2}=0.
\end{align}
Eq.~(\ref{eqnB20}) is identical to Eq.~(\ref{eqnB2}). Then we can take Eq.~(\ref{eqnB12}) into Eq.~(\ref{eqnB19}), and the difference from before is that we will get a second-order recursion relation, i.e.~$\alpha_{2\mathsf{k}+2}$ should be determined by $\alpha_{2\mathsf{k}}$ and $\alpha_{2\mathsf{k}-2}$,
\begin{align}
\label{eqnB21}
&\sum_{\mathsf{k}=-1}^{\mathsf{N}-1}\alpha_{2\mathsf{k}+2}[\frac{1}{2}(1+\frac{\eta}{2})(\ell+2\mathsf{k}+2)(\ell+2\mathsf{k}+1)\nonumber\\
&\!\ \!\ \!\ \!\ +(1+\frac{\eta}{2})(\ell+2\mathsf{k}+2)-\frac{1}{2}(1+\frac{\eta}{2})\ell(\ell+1)]r^{\ell+2\mathsf{k}}\nonumber\\
&\!\ \!\ +\sum_{\mathsf{k}=0}^{\mathsf{N}}\alpha_{2\mathsf{k}}[-\frac{1}{2}(1+\eta)(\ell+2\mathsf{k})(\ell+2\mathsf{k}-1)\nonumber\\
&\!\ \!\ \!\ \!\ \!\ \!\ -2(1+\eta)(\ell+2\mathsf{k})+\kappa+\frac{1}{2}(1+\eta)\ell(\ell+1)]r^{\ell+2\mathsf{k}}\nonumber\\
&\!\ \!\ +\sum_{\mathsf{k}=1}^{\mathsf{N}+1}\alpha_{2\mathsf{k}-2}[\frac{\eta}{4}(\ell+2\mathsf{k}-2)(\ell+2\mathsf{k}-3)\nonumber\\
&\!\ \!\ \!\ \!\ \!\ \!\ +\frac{3}{2}\eta(\ell+2\mathsf{k}-2)-\frac{\eta}{4}\ell(\ell+1)]r^{\ell+2\mathsf{k}}=0.
\end{align}
However, if $\mathsf{N}$ is finite, an equation for $\alpha_{2\mathsf{N}}r^{\ell+2\mathsf{N}+2}$ gives
\begin{align}
\label{eqnB22}
&\frac{1}{4}(\ell+2\mathsf{N})(\ell+2\mathsf{N}-1)+\frac{3}{2}(\ell+2\mathsf{N})-\frac{1}{4}\ell(\ell+1)\nonumber\\
=&(\mathsf{N}+1)\ell+\frac{1}{2}\mathsf{N}(2\mathsf{N}+5)=0,
\end{align}
whose only solution is $\mathsf{N}=\ell=0$, and an equation for $\alpha_{2\mathsf{N}}r^{\ell+2\mathsf{N}}$ gives $\kappa=0$. Besides this trivial zero mode ($\delta n$ is a constant), we should take $\mathsf{N}=\infty$.

When $\mathsf{N}=\infty$ and $\mathsf{k}\rightarrow\infty$, we can apply an expansion of $\mathsf{k}^{-1}$. At the leading order, we get a recursion relation,
\begin{align}
\label{eqnB23}
(2+\eta)\alpha_{2\mathsf{k}+2}-2(1+\eta)\alpha_{2\mathsf{k}}+\eta\alpha_{2\mathsf{k}-2}=0,
\end{align}
which leads to
\begin{align}
\label{eqnB24}
\alpha_{2\mathsf{k}}=A_1+A_2(\frac{\eta}{2+\eta})^{\mathsf{k}}.
\end{align}
So there are two parts of $\alpha_{2\mathsf{k}}$ with different decay exponents. Let us concentrate on the case of $\eta>-1$ so that $|\frac{\eta}{2+\eta}|<1$. At the next-to-leading order, similar to Eq.~(\ref{eqnB9}), $A_1$ and $A_2$ in Eq.~(\ref{eqnB24}) are no longer constants but proportional to powers of $\mathsf{k}$. Because Eq.~(\ref{eqnB23}) is linear, we can solve $A_1$ and $A_2$ separately, namely we can take only one of them to be nonzero. Taking $\alpha_{2\mathsf{k}}=\frac{C_1}{\mathsf{k}^{\zeta_1}}$, we get
\begin{align}
\label{eqnB25}
&(1-\frac{\zeta_1}{\mathsf{k}})(1+\frac{\eta}{2})(1+\frac{2\ell+5}{2\mathsf{k}})\nonumber\\
&-(1+\eta)(1+\frac{2\ell+3}{2\mathsf{k}})\nonumber\\
&+(1+\frac{\zeta_1}{\mathsf{k}})\frac{\eta}{2}(1+\frac{2\ell+1}{2\mathsf{k}})+\mathcal{O}(\frac{1}{\mathsf{k}^2})=0,
\end{align}
so the power $\zeta_1$ is determined,
\begin{align}
\label{eqnB26}
\zeta_1=(1+\frac{\eta}{2})(\ell+\frac{5}{2})-(1+\eta)(\ell+\frac{3}{2})+\eta(\frac{\ell}{2}+\frac{1}{4})=1.
\end{align}
Then we get $\alpha_{2\mathsf{k}}=\frac{C_1}{\mathsf{k}}+\mathcal{O}(\frac{1}{\mathsf{k}^2})$, which leads to the divergence shown in Eq.~(\ref{eqnB11}). To avoid the divergence, in the asymptotic behavior of $\alpha_{2\mathsf{k}}$ shown in Eq.~(\ref{eqnB24}), only the second part should be nonzero. Taking $\alpha_{2\mathsf{k}}=\frac{C_2}{\mathsf{k}^{\zeta_2}}(\frac{\eta}{2+\eta})^{\mathsf{k}}$, we get
\begin{align}
\label{eqnB27}
&(1-\frac{\zeta_2}{\mathsf{k}})(1+\frac{\eta}{2})(1+\frac{2\ell+5}{2\mathsf{k}})\eta^2\nonumber\\
&-(1+\eta)(1+\frac{2\ell+3}{2\mathsf{k}})\eta(2+\eta)\nonumber\\
&+(1+\frac{\zeta_2}{\mathsf{k}})\frac{\eta}{2}(1+\frac{2\ell+1}{2\mathsf{k}})(2+\eta)^2+\mathcal{O}(\frac{1}{\mathsf{k}^2})=0,
\end{align}
so the power $\zeta_2$ is determined,
\begin{align}
\label{eqnB28}
&\zeta_2=[(1+\frac{\eta}{2})(\ell+\frac{5}{2})\eta^2-(1+\eta)(\ell+\frac{3}{2})\eta(2+\eta)\nonumber\\
&\!\ \!\ +\eta(\frac{\ell}{2}+\frac{1}{4})(2+\eta)^2][-(1+\frac{\eta}{2})\eta^2+\frac{\eta}{2}(2+\eta)^2]^{-1}=1.
\end{align}
Then we get the asymptotic behavior of $\alpha_{2\mathsf{k}}$,
\begin{align}
\label{eqnB29}
\alpha_{2\mathsf{k}}=[\frac{C_2}{\mathsf{k}}+\mathcal{O}(\frac{1}{\mathsf{k}^2})](\frac{\eta}{2+\eta})^{\mathsf{k}}.
\end{align}

To make $C_1=0$, $\kappa$ should take discrete values. However, because $C_1=0$ is an asymptomatic condition for the sequence $\alpha_{2\mathsf{k}}$, we can only solve $\kappa$ numerically. From the recursion relation for $\mathsf{k}\geq 0$,
\begin{align}
\label{eqnB30}
&[\frac{1}{2}(1+\frac{\eta}{2})(\ell+2\mathsf{k}+2)(\ell+2\mathsf{k}+1)\nonumber\\
&\!\ \!\ \!\ \!\ +(1+\frac{\eta}{2})(\ell+2\mathsf{k}+2)-\frac{1}{2}(1+\frac{\eta}{2})\ell(\ell+1)]\alpha_{2\mathsf{k}+2}\nonumber\\
=&[\frac{1}{2}(1+\eta)(\ell+2\mathsf{k})(\ell+2\mathsf{k}-1)\nonumber\\
&\!\ \!\ \!\ \!\ +2(1+\eta)(\ell+2\mathsf{k})-\kappa-\frac{1}{2}(1+\eta)\ell(\ell+1)]\alpha_{2\mathsf{k}}\nonumber\\
&-[\frac{\eta}{4}(\ell+2\mathsf{k}-2)(\ell+2\mathsf{k}-3)\nonumber\\
&\!\ \!\ \!\ \!\ +\frac{3}{2}\eta(\ell+2\mathsf{k}-2)-\frac{\eta}{4}\ell(\ell+1)]\alpha_{2\mathsf{k}-2},
\end{align}
where $\alpha_{-2}\equiv 0$, when $\alpha_0$ is normalized to be $1$, we can get $\alpha_{2\mathsf{k}_0+2}(\kappa)$ and $\alpha_{2\mathsf{k}_0}(\kappa)$ for some large enough integer $\mathsf{k}_0$. Then $\kappa$ can be solved by (Eq.~(\ref{eqn41}))
\begin{align}
\label{eqnB31}
\alpha_{2\mathsf{k}_0+2}=\alpha_{2\mathsf{k}_0}\frac{\mathsf{k}_0}{\mathsf{k}_0+1}\frac{\eta}{2+\eta}.
\end{align}
Eq.~(\ref{eqnB31}) is an equation of order ($\mathsf{k}_0+1$), so we will get ($\mathsf{k}_0+1$) solutions of $\kappa$. When $k_0\rightarrow\infty$, we should get infinite solutions of $\kappa$, which form a discrete spectrum. Note that when $\eta=0$, Eq.~(\ref{eqnB31}) becomes the condition that $\mathsf{N}$ in Eq.~(\ref{eqnB12}) is finite, which is consistent with our previous result.

\subsubsection{\label{AppendixB22}2D case}

Consider Eq.~(\ref{eqn39}) with $\eta\neq 0$,
\begin{align}
\label{eqnB32}
&\frac{1}{2}[(1+\frac{\eta}{2})-(1+\eta)r^2+\frac{\eta}{2}r^4]\frac{d^2\rho}{d r^2}\nonumber\\
&\!\ \!\ +\frac{1}{2}[(1+\frac{\eta}{2})-3(1+\eta)r^2+\frac{5\eta}{2}r^4]\frac{1}{r}\frac{d\rho}{d r}\nonumber\\
&\!\ \!\ \!\ \!\ +\{\kappa r^2-\frac{1}{2}m^2[(1+\frac{\eta}{2})-(1+\eta)r^2+\frac{\eta}{2}r^4]\}\frac{\rho}{r^2}=0.
\end{align}
We still have Eq.~(\ref{eqnB14}). Taking $\mathsf{m}\geq 0$ for simplicity, we get a recursion relation,
\begin{align}
\label{eqnB33}
&[\frac{1}{2}(1+\frac{\eta}{2})(\mathsf{m}+2\mathsf{k}+2)(\mathsf{m}+2\mathsf{k}+1)\nonumber\\
&\!\ \!\ \!\ \!\ +\frac{1}{2}(1+\frac{\eta}{2})(\mathsf{m}+2\mathsf{k}+2)-\frac{1}{2}(1+\frac{\eta}{2})\mathsf{m}^2]\alpha_{2\mathsf{k}+2}\nonumber\\
&+[-\frac{1}{2}(1+\eta)(\mathsf{m}+2\mathsf{k})(\mathsf{m}+2\mathsf{k}-1)\nonumber\\
&\!\ \!\ \!\ \!\ -\frac{3}{2}(1+\eta)(\mathsf{m}+2\mathsf{k})+\kappa+\frac{1}{2}(1+\eta)\mathsf{m}^2]\alpha_{2\mathsf{k}}\nonumber\\
&+[\frac{\eta}{4}(\mathsf{m}+2\mathsf{k}-2)(\mathsf{m}+2\mathsf{k}-3)\nonumber\\
&\!\ \!\ \!\ \!\ +\frac{5}{4}\eta(\mathsf{m}+2\mathsf{k}-2)-\frac{\eta}{4}\mathsf{m}^2]\alpha_{2\mathsf{k}-2}=0.
\end{align}
Then $\mathsf{N}=\infty$ unless $\mathsf{m}=0$ and $\mathsf{N}=0$, because
\begin{align}
\label{eqnB34}
&\frac{1}{4}(\mathsf{m}+2\mathsf{N})(\mathsf{m}+2\mathsf{N}-1)+\frac{5}{4}(\mathsf{m}+2\mathsf{N})-\frac{1}{4}\mathsf{m}^2\nonumber\\
=&\mathsf{m}(2\mathsf{N}+1)+\mathsf{N}(\mathsf{N}+2)\geq 0.
\end{align}

When $\mathsf{N}=\infty$ and $\mathsf{k}\rightarrow\infty$, taking an expansion of $\mathsf{k}^{-1}$, at the leading order we get the same results as Eqs.~(\ref{eqnB23}, \ref{eqnB24}). Taking $\alpha_{2\mathsf{k}}=\frac{C_1}{\mathsf{k}^{\zeta_1}}$, we get
\begin{align}
\label{eqnB35}
&(1-\frac{\zeta_1}{\mathsf{k}})(1+\frac{\eta}{2})(1+\frac{\mathsf{m}+2}{\mathsf{k}})\nonumber\\
&-(1+\eta)(1+\frac{\mathsf{m}+1}{\mathsf{k}})\nonumber\\
&+(1+\frac{\zeta_1}{\mathsf{k}})\frac{\eta}{2}(1+\frac{\mathsf{m}}{\mathsf{k}})+\mathcal{O}(\frac{1}{\mathsf{k}^2})=0,
\end{align}
then
\begin{align}
\label{eqnB36}
\zeta_1=&(1+\frac{\eta}{2})(\mathsf{m}+2)\nonumber\\
&\!\ \!\ \!\ \!\ -(1+\eta)(\mathsf{m}+1)+\eta\frac{\mathsf{m}}{2}=1,
\end{align}
and we should take $C_1=0$ to make an eigenfunction convergent when $r\rightarrow 1$. Taking $\alpha_{2\mathsf{k}}=\frac{C_2}{\mathsf{k}^{\zeta_2}}(\frac{\eta}{2+\eta})^{\mathsf{k}}$, we get
\begin{align}
\label{eqnB37}
&(1-\frac{\zeta_2}{\mathsf{k}})(1+\frac{\eta}{2})(1+\frac{\mathsf{m}+2}{\mathsf{k}})\eta^2\nonumber\\
&-(1+\eta)(1+\frac{\mathsf{m}+1}{\mathsf{k}})\eta(2+\eta)\nonumber\\
&+(1+\frac{\zeta_2}{\mathsf{k}})\frac{\eta}{2}(1+\frac{\mathsf{m}}{\mathsf{k}})(2+\eta)^2+\mathcal{O}(\frac{1}{\mathsf{k}^2})=0,
\end{align}
so
\begin{align}
\label{eqnB38}
&\zeta_2=-[(1+\frac{\eta}{2})(\mathsf{m}+2)\eta^2-(1+\eta)(\mathsf{m}+1)\eta(2+\eta)\nonumber\\
&\!\ \!\ \!\ \!\ +\frac{\eta}{2}\mathsf{m}(2+\eta)^2][\eta^2+2\eta]^{-1}=1,
\end{align}
and Eq.~(\ref{eqnB31}) is still valid to determine eigenvalues numerically.

\subsection{\label{AppendixB3}Orthonormal relation}

The linear differential operators in Eqs.~(\ref{eqnB19}, \ref{eqnB32}) are non-Hermitian if we use the simplest definition of an inner product in a Hilbert space, so we need to apply the Sturm-Liouville theory~\cite{Mathews1971} to find proper definitions of the inner product. For an eigenfunction of a second-order linear differential operator,
\begin{align}
\label{eqnB39}
P(r)\frac{d^2\rho}{d r^2}+Q(r)\frac{d \rho}{d r}+[\kappa W(r)-S(r)]\rho=0,
\end{align}
it has a Sturm-Liouville form,
\begin{align}
\label{eqnB40}
\frac{d}{dr}[p(r)\frac{d\rho}{dr}]+[\kappa w(r)-s(r)]\rho=0,
\end{align}
where
\begin{align}
\label{eqnB41}
&p(r)=e^{\int\frac{Q(r)}{P(r)}dr},\quad w(r)=\frac{W(r)}{P(r)}e^{\int\frac{Q(r)}{P(r)}dr},\nonumber\\
&s(r)=\frac{S(r)}{P(r)}e^{\int\frac{Q(r)}{P(r)}dr}.
\end{align}
When a boundary condition is satisfied,
\begin{align}
\label{eqnB42}
p(r)[\rho_{\mathsf{n}_2,\ell}\frac{d\rho^*_{\mathsf{n}_1,\ell}}{dr}-\rho^*_{\mathsf{n}_1,\ell}\frac{d\rho_{\mathsf{n}_2,\ell}}{dr}]\big{|}_{r=0}^{r=1}=0,
\end{align}
the eigenfunctions can be normalized to form an orthonormal basis with an inner product,
\begin{align}
\label{eqnB43}
(\rho_{\mathsf{n}_1,\ell},\rho_{\mathsf{n}_2,\ell})\equiv\int_0^1 \frac{w(r)}{w}\rho_{\mathsf{n}_1,\ell}^*(r)\rho_{\mathsf{n}_2,\ell}(r)dr=\delta_{\mathsf{n}_1,\mathsf{n}_2},
\end{align}
where $w$ is an arbitrary coefficient.

\subsubsection{\label{AppendixB31}3D case}

In the 3D case,
\begin{align}
\label{eqnB44}
&p(r)=e^{\int \frac{Q(r)}{P(r)}dr}\nonumber\\
=&\mathrm{exp}\{\int\frac{(1+\frac{\eta}{2})-2(1+\eta)r^2+\frac{3}{2}\eta r^4}{\frac{r}{2}[(1+\frac{\eta}{2})-(1+\eta)r^2+\frac{\eta}{2}r^4]}dr\}\nonumber\\
=&r^2(1-r^2)(2+\eta-\eta r^2),
\end{align}
\begin{align}
\label{eqnB45}
w(r)=&\frac{2}{(1+\frac{\eta}{2})-(1+\eta)r^2+\frac{\eta}{2}r^4}r^2(1-r^2)(2+\eta-\eta r^2)\nonumber\\
=&4r^2,
\end{align}
\begin{align}
\label{eqnB46}
s(r)=\ell(\ell+1)(1-r^2)(2+\eta-\eta r^2).
\end{align}
Near $r=0$ we have got $\rho(r)\sim r^\ell$. Near $r=1$, we have required that $\rho(r)$ is finite, and by an expansion of $1-r$ we have $\frac{d\rho}{dr}=\kappa \rho$, so $\frac{d\rho}{dr}$ is also finite. Because $p(0)=p(1)=0$, for two different eigenfunctions $\rho_{\mathsf{n}_1,\ell}(r)$ and $\rho_{\mathsf{n}_2,\ell}(r)$, the condition Eq.~(\ref{eqnB42}) is satisfied. So we get the orthonormal relation in Eq.~(\ref{eqn42}),
\begin{align}
\label{eqnB47}
(\rho_{\mathsf{n}_1,\ell},\rho_{\mathsf{n}_2,\ell})\equiv\int_0^1 r^2\rho_{\mathsf{n}_1,\ell}^*(r)\rho_{\mathsf{n}_2}(r)dr=\delta_{\mathsf{n}_1,\mathsf{n}_2},
\end{align}
where we have taken $w=4$ in Eq.~(\ref{eqnB43}).

\subsubsection{\label{AppendixB32}2D case}

In the 2D case,
\begin{align}
\label{eqnB48}
p(r)=&\mathrm{exp}\{\int\frac{(1+\frac{\eta}{2})-3(1+\eta)r^2+\frac{5}{2}\eta r^4}{r[(1+\frac{\eta}{2})-(1+\eta)r^2+\frac{\eta}{2}r^4]}dr\}\nonumber\\
=&r(1-r^2)(2+\eta-\eta r^2),
\end{align}
\begin{align}
\label{eqnB49}
w(r)=4r,\quad s(r)=\frac{m^2}{r}(1-r^2)(2+\eta-\eta r^2).
\end{align}
Taking $w=4$, we get the orthonormal relation in Eq.~(\ref{eqn42}),
\begin{align}
\label{eqnB50}
(\rho_{\mathsf{n}_1,\mathsf{m}},\rho_{\mathsf{n}_2,\mathsf{m}})\equiv\int_0^1 r\rho_{\mathsf{n}_1,\mathsf{m}}^*(r)\rho_{\mathsf{n}_2,\mathsf{m}}(r)dr=\delta_{\mathsf{n}_1,\mathsf{n}_2}.
\end{align}

\section{\label{AppendixC}Discussions on tight-binding models}

In this Appendix, as a supplement to the continuous model Eqs.~(\ref{eqn1}--\ref{eqn3}), we discuss Bloch wavefunction effects in tight-binding models defined in discretized lattices. We first clarify a general tight-binding limit of the continuous model in Subsection \ref{AppendixC1}. We then focus on the simplest example, bipartite tight-binding models, and calculate their Bloch wavefunction quantities in Subsection \ref{AppendixC2}.

\subsection{\label{AppendixC1}General tight-binding limit}

In the tight-binding limit, the integral of $\bm{r}$ in a unit cell is substituted by a summation of sublattice indices times the volume of a unit cell. Namely, our Hamiltonian becomes
\begin{align}
\label{eqnC1}
&H_{0}=\Omega_{\mathrm{cell}}^2\sum_{\bm{R},\bm{R}'}\sum_{\sigma,\sigma'=0}^{\mathsf{n}_s-1}a^\dagger_{\bm{R}+\bm{t}_\sigma,\sigma}\nonumber\\
&\!\ \!\ \!\ \!\ \cdot(\bm{H}_{0,\mathrm{eff}})_{\sigma,\sigma'}(\bm{R}-\bm{R}'+\bm{t}_\sigma-\bm{t}_{\sigma'})a_{\bm{R}'+\bm{t}_{\sigma'},\sigma'}\nonumber\\
&\!\ \!\ +\frac{g_{\mathrm{eff}}}{2}\Omega_{\mathrm{cell}}\sum_{\bm{R}}\sum_{\sigma=0}^{\mathsf{n}_s-1}a^\dagger_{\bm{R}+\bm{t}_\sigma,\sigma}a^\dagger_{\bm{R}+\bm{t}_\sigma,\sigma}a_{\bm{R}+\bm{t}_\sigma,\sigma}a_{\bm{R}+\bm{t}_\sigma,\sigma}\nonumber\\
&\!\ \!\ +\Omega_{\mathrm{cell}}\sum_{\bm{R}}\sum_{\sigma=0}^{\mathsf{n}_s-1}U(\bm{R}+\bm{t}_\sigma)a^\dagger_{\bm{R}+\bm{t}_\sigma,\sigma}a_{\bm{R}+\bm{t}_\sigma,\sigma},
\end{align}
where $\sigma$ and $\sigma'$ are sublattice indices, $\bm{R}$ and $\bm{R}'$ take lattice vectors, $\mathsf{n}_s$ is the number of sublattices in a unit cell, $\bm{t}_\sigma$ is the relative position of the $\sigma$-th sublattice in a unit cell. Different from a more commonly used convention that different sublattices in the same unit cell are regarded as in the same position, here we keep $\bm{t}_\sigma$ in Fourier transforms to make crystal symmetries more explicit. This convention results in discontinuity of a $\bm{k}$-space Hamiltonian at the boundary of the first Brillouin zone, but this is not a problem for us because we only care about the Bloch states near the band bottom.

In Eq.~(\ref{eqnC1}), $U(\bm{R}+\bm{t}_\sigma)\equiv U(\bm{r}=\bm{R}+\bm{t}_\sigma)$ and its Fourier transform are defined in the continuous space as the general cases of the continuous model,
\begin{align}
\label{eqnC2}
U(\bm{R}+\bm{t}_\sigma)=&\int_\infty\frac{d^d\bm{k}}{(2\pi)^d}e^{i\bm{k}\cdot(\bm{R}+\bm{t}_\sigma)}U_{\bm{k}}\nonumber\\
\approx&\int_{\mathrm{BZ}}\frac{d^d\bm{k}}{(2\pi)^d}e^{i\bm{k}\cdot(\bm{R}+\bm{t}_\sigma)}U_{\bm{k}}.
\end{align}
Because now $a_\sigma(\bm{R}+\bm{t}_{\sigma})$ and $(H_{0,\mathrm{eff}})_{\sigma,\sigma'}(\bm{R}-\bm{R}'+\bm{t}_{\sigma}-\bm{t}_{\sigma'})$ are defined in the lattice, to recover the general result of the low-energy effective action in the Bloch basis, their Fourier transforms are defined by
\begin{align}
\label{eqnC3}
a_{\bm{R}+\bm{t}_\sigma,\sigma}=\sum_{\mathsf{n}=0}^{\mathsf{n}_s-1}\int_{\mathrm{BZ}}\frac{d^d\bm{k}}{(2\pi)^{\frac{d}{2}}}e^{i\bm{k}\cdot(\bm{R}+\bm{t}_\sigma)}u_{\mathsf{n},\bm{k},\sigma}b_{\mathsf{n},\bm{k}},
\end{align}
\begin{align}
\label{eqnC4}
&(\bm{H}_{0,\mathrm{eff}})_{\sigma,\sigma'}(\bm{R}-\bm{R}'+\bm{t}_\sigma-\bm{t}_{\sigma'})\nonumber\\
=&\sum_{\mathsf{n},\mathsf{n}'=0}^{\mathsf{n}_s-1}\int_{\mathrm{BZ}}\frac{d^d\bm{k}}{(2\pi)^{d}}e^{i\bm{k}\cdot(\bm{R}-\bm{R}'+\bm{t}_\sigma-\bm{t}_{\sigma'})}(\bm{H}_{0,\mathrm{eff}})_{\sigma,\sigma',\bm{k}}\nonumber\\
=&\sum_{\mathsf{n},\mathsf{n}'=0}^{\mathsf{n}_s-1}\int_{\mathrm{BZ}}\frac{d^d\bm{k}}{(2\pi)^d}e^{i\bm{k}\cdot(\bm{R}-\bm{R}'+\bm{t}_\sigma-\bm{t}_{\sigma'})}u^*_{\mathsf{n},\bm{k},\sigma}u_{\mathsf{n},\bm{k},\sigma'}(\bm{H}_{0,\mathrm{eff}})_{\mathsf{n},\bm{k}},
\end{align}
where $\mathsf{n}$ is the band index, $u_{\mathsf{n},\bm{k},\sigma}$ is the normalized eigenvector of $(\bm{H}_{0,\mathrm{eff}})_{\sigma,\sigma',\bm{k}}$ in the sublattice space,
\begin{align}
\label{eqnC5}
\sum_{\sigma=0}^{\mathsf{n}_s-1} u^*_{\mathsf{n},\bm{k},\sigma}u_{\mathsf{n}',\bm{k},\sigma}=\delta_{\mathsf{n},\mathsf{n}'}.
\end{align}

Note that the interaction strength $g_{\mathrm{eff}}$ in Eq.~(\ref{eqnC1}) is different from $g$ in Eq.~(\ref{eqn3}), so we add a subscript ``$\mathrm{eff}$" to emphasize this, although they play the same role for a low-energy effective theory. The tight-binding limit makes $M\rightarrow\infty$ and $g_{\mathrm{eff}}=\mathcal{O}(gM)\rightarrow\infty$. This is because atoms are constrained at only several points in the unit cell, which significantly enlarges the effective repulsive interaction. Though, as there must be finite widths of Wannier functions in reality, we can renormalize $g_{\mathrm{eff}}$ to be finite. 

A Bloch wavefunction quantity $M_{\mathsf{n},\mathrm{eff}}$ is defined by
\begin{align}
\label{eqnC6}
M_{\mathsf{n},\mathrm{eff}}(\bm{k})=\mathsf{n}_s\sum_{\sigma=0}^{\mathsf{n}_s-1} u^*_{\mathsf{n},\bm{k},\sigma} u^*_{\mathsf{n},\bm{k},\sigma} u_{\mathsf{n},\bm{k},\sigma} u_{\mathsf{n},\bm{k},\sigma},
\end{align}
and other Bloch wavefunction quantities are defined similarly. When the sublattices become equivalent, $|u_{\mathsf{n},\bm{k},\sigma}|=\frac{1}{\sqrt{\mathsf{n}_s}}$ and $M_{\mathsf{n},\mathrm{eff}}(\bm{k})=1$; for a general case, $|u_{\mathsf{n},\bm{k},\sigma}|$ depends on $\sigma$ and $M_{\mathsf{n},\mathrm{eff}}(\bm{k})>1$. We focus on the lowest band, i.e.~$\mathsf{n}=0$, whose dispersion is assumed to be $(\bm{H}_{0,\mathrm{eff}})_{0,\bm{k}}=\frac{k^2}{2m}+\mathcal{O}(k^3)$ near the band bottom $\bm{k}=\bm{0}$. 

In the next subsection, we only consider the tight-binding limit and only discuss the quantities with the subscript ``eff", so we will omit the subscript ``eff" for convenience. We also omit the band index in $(\bm{H}_0)_{\bm{k}}$ when denoting a matrix and in $u_{\bm{k},\sigma}$ when denoting the lowest band.

\subsection{\label{AppendixC2}Bloch wavefunction quantities of bipartite tight-binding models}

Next, we focus on bipartite lattices ($\mathsf{n}_s=2$). For a 2D lattice, suppose there is a $\mathbb{Z}_n$ ($n\geq 3$) rotational symmetry; for a 3D lattice, suppose it belongs to the cubic crystal system. Near the band bottom $\bm{k}=\bm{0}$, in the sublattice space we can expand the single-body Hamiltonian by Pauli matrices,
\begin{align}
\label{eqnC7}
(\bm{H}_0)_{\bm{k}}=&\tilde{h}_{\bm{k}}\mathrm{sin}\Theta_{\bm{k}}\mathrm{cos}\Phi_{\bm{k}}\bm{\sigma}_x+\tilde{h}_{\bm{k}}\mathrm{sin}\Theta_{\bm{k}}\mathrm{sin}\Phi_{\bm{k}}\bm{\sigma}_y\nonumber\\
&+\tilde{h}_{\bm{k}}\mathrm{cos}\Theta_{\bm{k}}\bm{\sigma}_z+\bar{h}_{\bm{k}}\bm{\sigma}_0,
\end{align}
where we neglect a constant which shifts the energy of $\bm{k}=\bm{0}$ of the lower band to be zero. Then the lower-band eigenvector is given by
\begin{align}
\label{eqnC8}
\bm{u}_{\bm{k}}\equiv\left(\begin{array}{c}u_{\bm{k},1}\\u_{\bm{k},2}\end{array}\right)=\left(\begin{array}{c}-\mathrm{sin}\frac{\Theta_{\bm{k}}}{2}\\ \mathrm{cos}\frac{\Theta_{\bm{k}}}{2}e^{i\Phi_{\bm{k}}}\end{array}\right).
\end{align}
The Hamiltonian near $\bm{k}=\bm{0}$ is constrained by the lattice symmetry,
\begin{align}
\label{eqnC9}
(\bm{H}_0)_{\bm{k}}=&(a_x-\frac{1}{2}b_xk^2)\bm{\sigma}_x+(a_y-\frac{1}{2}b_yk^2)\bm{\sigma}_y\nonumber\\
&+(a_z-\frac{1}{2}b_zk^2)\bm{\sigma}_z+\frac{1}{2}b_0k^2\bm{\sigma}_0+\mathcal{O}(k^3).
\end{align}
For simplicity, we take $a_{x,y,z}\geq 0$, $b_{x,y,z,0}\geq 0$, and it can be generalized to other cases. 

If there is a time-reversal symmetry, i.e.~$(\bm{H}_0)^*_{\bm{k}}=(\bm{H}_0)_{-\bm{k}}$, then $a_y=b_y=0$. If there is an inversion symmetry that exchanges two sublattices (e.g. 2D honeycomb lattice), i.e.~$\bm{\sigma}_x(\bm{H}_0)_{\bm{k}}\bm{\sigma}_x=(\bm{H}_0)_{-\bm{k}}$, then $a_y=b_y=a_z=b_z=0$. If there is an inversion symmetry that keeps the sublattices invariant (e.g. 2D bipartite square lattice), i.e.~$(\bm{H}_0)_{\bm{k}}=(\bm{H}_0)_{-\bm{k}}$, it does not give additional constraints to Eq.~(\ref{eqnC9}). Besides, we take the permutation of sublattices as our definition of the sublattice symmetry, i.e.~$\bm{\sigma}_x(\bm{H}_0)_{\bm{k}}\bm{\sigma}_x=(\bm{H}_0)_{\bm{k}}$, which is different from a conventional one, i.e.~$\bm{\sigma}_z(\bm{H}_0)_{\bm{k}}\bm{\sigma}_z=-(\bm{H}_0)_{\bm{k}}$. The sublattice symmetry requires $a_y=b_y=a_z=b_z=0$. 

We can calculate the Bloch wavefunction quantities near $\bm{k}=\bm{0}$, which are much simplified by $\partial_{k_x}\Theta_{\bm{0}}=\partial_{k_x}\Phi_{\bm{0}}=0$,
\begin{align}
\label{eqnC10}
&M_d(\bm{k})=M(\bm{k})-1\nonumber\\
=&2(\mathrm{sin}^4\frac{\Theta_{\bm{k}}}{2}+\mathrm{cos}^4\frac{\Theta_{\bm{k}}}{2})-1=\mathrm{cos}^2\Theta_{\bm{k}},
\end{align}
\begin{align}
\label{eqnC11}
\partial_{k_x}M(\bm{k})=-\mathrm{sin}2\Theta_{\bm{k}}\partial_{k_x}\Theta_{\bm{k}},
\end{align}
\begin{align}
\label{eqnC12}
M''(\bm{0})=\partial_{k_x}^2M(\bm{0})=-\mathrm{sin}2\Theta_{\bm{0}}\partial_{k_x}^2\Theta_{\bm{0}},
\end{align}
\begin{align}
\label{eqnC13}
\partial_{k_x}\bm{u}_{\bm{k}}=\left(\begin{array}{c}-\frac{1}{2}(\partial_{k_x}\Theta_{\bm{k}})\mathrm{cos}\frac{\Theta_{\bm{k}}}{2}\\
-\frac{1}{2}(\partial_{k_x}\Theta_{\bm{k}})\mathrm{sin}\frac{\Theta_{\bm{k}}}{2}e^{i\Phi_{\bm{k}}}+i(\partial_{k_x}\Phi_{\bm{k}})\mathrm{cos}\frac{\Theta_{\bm{k}}}{2}e^{i\Phi_{\bm{k}}}\end{array}\right),
\end{align}
\begin{align}
\label{eqnC14}
\partial_{k_x}^2\bm{u}_{\bm{0}}=\left(\begin{array}{c}-\frac{1}{2}(\partial_{k_x}^2\Theta_{\bm{0}})\mathrm{cos}\frac{\Theta_{\bm{0}}}{2}\\
-\frac{1}{2}(\partial_{k_x}^2\Theta_{\bm{0}})\mathrm{sin}\frac{\Theta_{\bm{0}}}{2}e^{i\Phi_{\bm{0}}}+i(\partial_{k_x}^2\Phi_{\bm{0}})\mathrm{cos}\frac{\Theta_{\bm{0}}}{2}e^{i\Phi_{\bm{0}}}\end{array}\right),
\end{align}
\begin{align}
\label{eqnC15}
S(\bm{0})=&\mathrm{Im}[2\sum_{\sigma=0}^{1} u^*_{\bm{0},\sigma} u^*_{\bm{0},\sigma} u_{\bm{0},\sigma} \partial_{k_x^2}u_{\bm{0},\sigma}\nonumber\\
&\!\ \!\ \!\ \!\ -M(\bm{0})\sum_{\sigma=0}^{1}u^*_{\bm{0},\sigma}\partial_{k_x^2}u_{\bm{0},\sigma}]\nonumber\\
=&2\mathrm{cos}^4\frac{\Theta_{\bm{0}}}{2}\partial_{k_x}^2\Phi_{\bm{0}}-(1+\mathrm{cos}^2\Theta_{\bm{0}})\mathrm{cos}^2\frac{\Theta_{\bm{0}}}{2}\partial_{k_x}^2\Phi_{\bm{0}}\nonumber\\
=&\frac{1}{4}\mathrm{sin}2\Theta_{\bm{0}}\partial_{k_x}^2\Phi_{\bm{0}}.
\end{align}
When there is the time-reversal symmetry, we get $\Phi_{\bm{k}}=0$, so $S(\bm{k})=0$. When there is the sublattice symmetry or a 3D inversion symmetry that exchanges the two sublattices, we get $\mathrm{cos}\Theta_{\bm{0}}=0$, so $M_d(\bm{0})=M''(\bm{0})=S(\bm{0})=0$, where there are no lattice-induced wavefunction effects in the low-energy effective theory.

When the sublattice symmetry and the time-reversal symmetry are weakly broken, $\frac{a_y}{a_x}$, $\frac{b_y}{b_x}$, $\frac{a_z}{a_x}$, $\frac{b_z}{b_x}$ are small. Then at the leading order, we get
\begin{align}
\label{eqnC16}
m\approx \frac{1}{b_0+b_x},\quad M_d(\bm{0})\approx\frac{a_z^2}{a_x^2},
\end{align}
\begin{align}
\label{eqnC17}
\Theta_{\bm{k}}=&\frac{\pi}{2}-\mathrm{arctan}\frac{a_z-\frac{1}{2}b_zk^2}{\sqrt{(a_x-\frac{1}{2}b_xk^2)^2+(a_y-\frac{1}{2}b_yk^2)^2}},\nonumber\\
\Phi_{\bm{k}}=&\mathrm{arctan}\frac{a_y-\frac{1}{2}b_yk^2}{a_x-\frac{1}{2}b_xk^2},
\end{align}
\begin{align}
\label{eqnC18}
&\partial_{k_x}^2\Theta_{\bm{0}}=2\frac{\partial\Theta_{\bm{0}}}{\partial k^2}=\frac{1}{1+\frac{a_z^2}{a_x^2+a_y^2}}[\frac{b_z}{(a_x^2+a_y^2)^{\frac{1}{2}}}\nonumber\\
&\!\ \!\ \!\ \!\ -\frac{1}{2}\frac{2a_z(a_xb_x+a_yb_y)}{(a_x^2+a_y^2)^{\frac{3}{2}}}]\approx\frac{b_z}{a_x}-\frac{b_xa_z}{a_x^2},
\end{align}
\begin{align}
\label{eqnC19}
\partial_{k_x}^2\Phi_{\bm{0}}=&2\frac{\partial\Phi_{\bm{0}}}{\partial k^2}=-\frac{1}{1+\frac{a_y^2}{a_x^2}}\frac{a_xb_y-a_yb_x}{a_x^2}\nonumber\\
\approx&-\frac{b_y}{a_x}+\frac{b_xa_y}{a_x^2},
\end{align}
\begin{align}
\label{eqnC20}
M''(\bm{0})\approx&-\frac{2a_z}{a_x}(\frac{b_z}{a_x}-\frac{b_xa_z}{a_x^2}),\nonumber\\
S(\bm{0})\approx&-\frac{a_z}{2a_x}(\frac{b_y}{a_x}-\frac{b_xa_y}{a_x^2}).
\end{align}
If $\frac{a_y}{a_x}$, $\frac{b_y}{b_x}$, $\frac{a_z}{a_x}$, $\frac{b_z}{b_x}$ are in the same order, $\mu$ and $a_x$ (or a bandwidth) are in the same order, then without fine-tuning, $M_d(\bm{0})$, $m\mu M''(\bm{0})$, and $m\mu S(\bm{0})$ are in the same order.

For the example given in Eq.~(\ref{eqn61}), from Eq.~(\ref{eqn62}) we get
\begin{align}
\label{eqnC21}
&b_0=2t_1a_0^2,\quad a_x=-t_2-4t_3\mathrm{cos}\theta_3,\nonumber\\
&b_x=-2t_3a_0^2\mathrm{cos}\theta_3,
\end{align}
\begin{align}
\label{eqnC22}
&a_y=-4t_3\mathrm{sin}\theta_3,\quad b_y=-2t_3a_0^2\mathrm{sin}\theta_3,\nonumber\\
&a_z=\Delta,\quad b_z=0.
\end{align}
We can apply a gauge transformation to take an additional sign for $a_{x,y}$ and $b_{x,y}$ so that they become positive. The gauge transformation is in the sublattice space instead of the momentum space, which does not affect the Bloch wavefunction quantities. When $\Delta$ and $\theta_3$ are small, taking Eqs.~(\ref{eqnC21}, \ref{eqnC22}) into Eqs.~(\ref{eqnC16}, \ref{eqnC20}) and keeping the leading order, we get the results of Eq.~(\ref{eqn63}), where two terms in $S(\bm{0})$ have been combined,
\begin{align}
\label{eqnC23}
S(\bm{0})=&-\frac{\Delta}{2(t_2+4t_3)}[\frac{2t_3a_0^2\theta_3}{t_2+4t_3}-\frac{8t_3^2a_0^2\theta_3}{(t_2+4t_3)^2}]\nonumber\\
=&-\frac{a_0^2t_2t_3\theta_3\Delta}{(t_2+4t_3)^3}.
\end{align}


\bibliography{Geometry_Superfluid}

\begin{thebibliography}{87}
\bibitem [{SM2()}]{SM2}%
  \BibitemOpen
  \href@noop {} {}\bibinfo {note} {When $-2<\eta\leq-1$, the series diverges at
  $r=1$. However, we cannot conclude that there is no solution to the
  differential equation. Further study is needed for whether the ground state
  is stable and whether there is a discrete dispersion of collective modes in
  this case.}\BibitemShut {Stop}%
\bibitem [{SM3()}]{SM3}%
  \BibitemOpen
  \href@noop {} {}\bibinfo {note} {An effect of $W$ can be shown in a
  higher-order contribution to the density distribution of the ground state in
  the external potential.}\BibitemShut {Stop}%
\end{thebibliography}

\end{document}